\documentclass{aa}
\usepackage[utf8]{inputenc} 

\usepackage{afterpage}
\usepackage{graphicx}
\usepackage{amsmath,amssymb} 
\usepackage{txfonts}
\usepackage[normalem]{ulem}
\usepackage[T1]{fontenc}
\usepackage{afterpage}
\usepackage{longtable}
\usepackage{lscape}
\usepackage{pdflscape}
\usepackage{amssymb}
\usepackage{natbib} 
\usepackage{wrapfig}
\usepackage{graphicx}
\usepackage{soul}
\usepackage{xcolor,color, colortbl}
\newcommand{\urlNewWindow}[1]{\href[pdfnewwindow=true]{#1}{\nolinkurl{#1}}}
\bibpunct{(}{)}{;}{a}{}{,}
\newcommand{\Ms}{~$M_\odot$\ }
\newcommand{\Rs}{\ensuremath{R_\odot}}

\newcommand{\ddr}[1]{\frac{\partial{}#1}{\partial{}r}}

 \def\lesssim{\mathrel{\hbox{\rlap{\hbox{\lower4pt\hbox{$\sim$}}}\hbox{$<$}}}}
 \def\gtrsim{\mathrel{\hbox{\rlap{\hbox{\lower4pt\hbox{$\sim$}}}\hbox{$>$}}}}
\newcommand{\eg}{{\it e.g. }}
\newcommand{\ie}{{\it i.e. }}
\definecolor{Gray}{gray}{0.9}

\usepackage{epstopdf}
\usepackage[]{sidecap}

\authorrunning{Amard et al.}
\titlerunning{Grid of rotating low-mass PMS stars}
\begin{document} 

   \title{First grids of low-mass stellar models and isochrones with self-consistent treatment of rotation}

   \subtitle{From 0.2 to 1.5\Ms at 7 metallicities from PMS to TAMS}

   \author{L. Amard
          \inst{1,2,3}
          \and
          A. Palacios\inst{2}
          \and 
          C. Charbonnel\inst{1,4}
          \and 
          F. Gallet\inst{5,1}
          \and
          C. Georgy\inst{1}
          \and 
          N. Lagarde\inst{6}
          \and
          L. Siess\inst{7}
          }
   \offprints{L. Amard: l.amard AT exeter.ac.uk}

   \institute{Department of Astronomy - University of Geneva - Chemin des Maillettes, 51 - CH-1290 Versoix, Switzerland 
         \and LUPM UMR 5299 CNRS/UM, Universit\'e de Montpellier, CC 72, 34095 Montpellier Cedex 05, France
         \and University of Exeter, Department of Physics \& Astronomy, Stoker Road, Devon, Exeter, EX4 4QL, UK
         \and IRAP, UMR 5277 CNRS and Universit\'e de Toulouse, 14 Av. E. Belin, F-31400 Toulouse, France
         \and Univ. Grenoble Alpes, CNRS, IPAG, 38000 Grenoble, France
         \and Institut UTINAM, CNRS UMR 6213, Univ. Bourgogne Franche-Comt\'e, OSU THETA Franche-Comt\'e-Bourgogne, Observatoire de Besan\c con, BP 1615, 25010, Besan\c con Cedex, France 
         \and Institut d'Astronomie et d'Astrophysique, Universit\'e Libre de Bruxelles (ULB), CP226, Boulevard du Triomphe, B-1050
Brussels, Belgium 
              }

   \date{Received September 15, 2996; accepted March 16, 2997}

  \abstract
   {}
   {We present an extended grid of state-of-the art stellar models for low-mass stars including updated physics (nuclear reaction rates, surface boundary condition, mass-loss rate, angular momentum transport, torque and rotation-induced mixing prescriptions). 
   We aim at evaluating the impact of wind braking, realistic atmospheric treatment, rotation and rotation-induced mixing on the structural and rotational evolution from the pre-main sequence to the turn-off. }
   {Using the STAREVOL code, we provide an updated PMS grid. We compute stellar models for 7 different metallicities, from  [Fe/H] = -1 dex to  [Fe/H] = +0.3 dex with a solar composition corresponding to $Z=0.0134$. The initial stellar mass ranges from 0.2 to 1.5\Ms with extra grid refinement around one solar mass. We also provide rotating models for three different initial rotation rates (slow, median and fast) with prescriptions for the wind braking and disc-coupling timescale calibrated on observed properties of young open clusters. The rotational mixing includes an up-to-date description of the turbulence anisotropy in stably stratified regions.}
   {The overall behaviour of our models at solar metallicity -- and its constitutive physics --  is validated through a detailed comparison with a variety of distributed evolutionary tracks. The main differences arise from the choice of surface boundary conditions and initial solar composition.
   The models including rotation with our prescription for angular momentum extraction and self-consistent formalism for angular momentum transport are able to reproduce the rotation period distribution observed in young open clusters over a broad mass-range. 
   These models are publicly available and may be used to analyse data coming from present and forthcoming asteroseismic and spectroscopic surveys such as Gaia, TESS and PLATO.}
   {}


   \maketitle
%

\section{Introduction}\label{sec:intro}
Along with mass and chemical composition, the angular momentum (hereafter AM) content is one of the fundamental characteristics of single stars \citep[see the review by][]{Maederbook}.
Rotation affects the whole stellar evolution from birth to death, with direct effects on the structure \citep[e.g.][]{ES76,MaederMeynet2001,Roxburgh2004,Rieutord2006}, mass-loss rate \citep[e.g.][]{OwockiGayley1997,Langer1998,MaederMeynetARAA2000,Georgyetal2011}, evolutionary path in the Hertzsprung-Russell diagram, asteroseismic properties  \citep[e.g.][]{Ballot2006,Eggenberger2010a,Lagarde2012,Reese2013,Bouabid2013,Prat2017},  lifetime, and on the internal and surface chemical composition of stars. 
It is also of crucial importance for potential life-hosting stellar systems because of the role played by rotation in generating magnetic fields by dynamo action \citep[e.g.][]{Noyesetal84,Wright2011,Vidotto2014,Johnstone2015c,Gallet2017,BrunBrowning2017}.

The case of low-mass stars is particularly interesting because their AM evolution involves various processes at different stages of their life. It starts with the collapse of the initial proto-stellar cloud in which jets and outflows remove of the order of 99\% of the AM content on a dynamical time-scale \citep[see \eg][]{Mathieu2004}. 
Then magnetic interactions between the star and its surrounding disk and latter between the stellar wind and the star's magnetic field determine the star's rotation velocity  on the main sequence. 

During the last decade,  stellar evolution models have strongly benefited from high quality photometry data from space missions, such as Kepler \citep[see \emph{e.g.}][]{Borucki2010,Gilliland2010}, and ground-based long term monitoring surveys \citep[see \emph{e.g.}][for a fairly complete overview]{BouvierPPVI}. 
By providing accurate surface rotation periods for large samples of low-mass stars, and  internal rotation profiles at some specific evolution phases by seismic analysis, these complementary observations improved our knowledge of the physical processes driving the rotational evolution of stars.

Over the same period, special care was brought to the development of new models for AM losses due to magnetized winds in low-mass \citep[see \emph{e.g.}][]{Pinto2011,RM12,Matt2012,VSP2013,Mattetal2015,Johnstone2015a,Reville2016,PantolmosMatt2017,Garraffo2018} and massive stars \citep[see \emph{e.g.}][]{Uddoula2009,Lauetal11}. Most of these prescriptions have been tested through post-processing computations of AM evolution based on pre-computed standard evolutionary tracks \citep[see \emph{e.g.}][]{GB13,GB15,Johnstone2015b,SadeghiA2017}, hence ignoring the effects of rotation on the structure and the evolution of the star. This  approach provides information on the magnitude of the torque and the degree of core-envelope {(de-)coupling} at the different phases of the evolution, and the results are compatible with the observed rotation period distribution of stars in clusters. 
However, it does not probe the actual physical processes that transport AM in the interior.

To reach a more consistent picture over a broader range of mass and chemical composition, we implemented these magnetic wind braking models directly into our stellar evolution code that can self-consistently treat rotation-induced transport processes (\textit{e.g.}, meridional circulation and turbulence).
We first applied this approach to solar-type stars in \cite{Amard2016}. In that study, we searched for the best combinations of prescriptions for internal transport of AM and surface braking by magnetized stellar winds, to account for the observed rotational periods in open clusters of different ages.
 We showed that the rotation period distributions can be successfully reproduced by models maintaining a certain amount of internal differential rotation even at late ages. 
We also confirmed that evolutionary models that only include AM transport by meridional circulation and turbulence, do not properly account for the rotation profile inside the Sun \citep{TurckChieze2010,Marques2013}, and the core rotation rates in subgiant and red giant stars  \citep{Eggenbergeretal12,Cellieretal2012}. 
These models also failed to reproduce the surface lithium abundances of solar-mass stars in young clusters \citep[\eg][]{SestitoRandich2005,TalonCharbonnel2010,SomersStassun2017}. As of today, the consensus is that additional  processes, like internal gravity waves or magnetic fields \citep{CharbonnelTalon2005Science,Eggenberger2005,TalonCharbonnel2008,Charbonnel2013,Li2014,Cantiello2014,Fuller2014,Belkacem2015,Jouve2015,Pincon2017}  might play an important role.

However, we have not yet fully investigated the complexity of rotation-driven hydrodynamical instabilities \citep[see \eg][]{Mathis2018,Jermyn2018}.
Recently we proposed an up-to-date description of anisotropic turbulence in stellar radiative regions \citep{Mathis2018} that induces a more efficient transport of AM than previous prescriptions. This prescription cannot yet reproduce the solid-body rotation profile at the age of the Sun but 
links for the first time the anisotropy of the turbulent transport in radiation zones to their stratification and rotation. This is a major improvement, which deserves a deeper investigation over a broad range of stellar masses and metallicities.

In the present grid of stellar models, we take into account up-to-date prescriptions for AM extraction by magnetized winds and AM transport by anisotropic turbulence. Our computations also include state-of-the-art model atmospheres and updated nuclear reaction rates. 
They cover the evolution from the PMS to the main sequence turnoff for stars with masses between 0.2 to 1.5\Ms and we consider seven metallicities ([Fe/H] between -1 dex and +0.3 dex). For each mass-metallicity combination, we compute one non-rotating (so-called standard) model, and three rotating models with different initial rotation rates (slow, median and fast) and disc lifetimes to cover the dispersion of rotation periods observed for stars of different masses and ages.
These stellar tracks are made available to the community, and we also provide the corresponding isochrones.

Despite the existence of other grids of rotating stellar models with various initial metallicities \citep[e.g.][]{Lagarde2012,Ekstrom2012,Yang2013,Georgy2013,Choietal2016}, this work presents for the first time a discussion of the effects of varying both the initial mass and metallicity on the transport of AM in low-mass stars undergoing magnetic braking.

The paper is structured as follows. In the next section, we describe the updated version of the STAREVOL code and present the various prescriptions used for input physics and our initial conditions. \S~\ref{sec:web} describes the content of the online material and in \S~\ref{sec:comp}, we compare our standard $Z_\odot$ tracks to other models computed with different codes. In \S~\ref{sec:AM}, we discuss the evolution of AM and compare the predictions of the solar metallicity grid to observed rotation period distributions in open clusters. Finally we briefly conclude in \S~\ref{sec:conclusion}. 


\section{The stellar evolution code : STAREVOL v3.40}\label{sec:code}
The models presented hereafter were computed with the stellar evolution code STAREVOL. The widely used PMS grid by \cite{Siess2000} was already computed with an early version of this code, as were the grids of low- and intermediate-mass stars by \cite{FC97}, \cite{Siess2002}, \cite{Lagarde2012}, \cite{Chantereauetal15} and SAGB stars \citep{Siess2007,Siess2010}. Here we use the latest version of the code (v3.40) jointly developed at Geneva and Montpellier Universities, which is an update of version v3.30 used in \cite{Amard2016}.  
We describe below the input prescriptions for the micro- and macro-physics of STAREVOL v3.40; in some cases we comment on the differences with respect to other grids from the literature (see also \S~\ref{sec:comp}). 

\subsection{Initial abundances and opacities}
\label{Sect:initabund}
We adopt the heavy elements mixture of \citet{AsplundGrevesse2009}, giving a reference value of solar photospheric metallicity $Z=0.013446$. A calibration of the solar model with the present input physics leads to an initial helium mass fraction $Y=0.2691$. We use the corresponding constant slope $\Delta Y/\Delta Z=1.60$ (with the primordial abundance $Y_0=0.2463$ based on WMAP-SBBN by \citealt{Cocetal04}) to set the initial helium mass fraction at a given  metallicity $Z$. We account for $\alpha-$elements enrichments below [Fe/H] $\leq$ -0.3 dex following the Galactic chemical evolution trends by \eg \cite{Fuhrmann2011}. The values are shown in Table~\ref{tab:abund}. Compared to the \citet{GN93} solar mixture used in \citet{Siess2000}, the solar abundances of almost all elements heavier than helium, and especially carbon, nitrogen, and oxygen, are significantly lower. Our initial chemical composition is slightly different from \cite{Lagarde2012} who also used \citet{AsplundGrevesse2009} heavy element mixture, because our solar calibration with updated physics leads to higher solar helium mass fraction ($Y=0.2691$ instead of 0.266) and helium to metals slope ($\Delta Y/\Delta Z=1.60$ instead of $1.29$).

Below 8000~K we use the  \cite{F05} opacities and above this temperature, the OPAL tables \cite{IglesiasRogers96}.
We use the same equation of state as  described in \cite{Siess2000}.

\begin{table}[h]
\begin{center}
\caption{Chemical abundances and metallicities - Scaled according to the solar chemical mixture by \citet{AsplundGrevesse2009}}
\begin{tabular}{ c c c c c }
\hline
\hline
[Fe/H] &  [$\alpha$/Fe] &    $Z$     & $Y$ \\
\hline
+0.3 & 0.0 & 0.02565 & 0.2884 \\ 
+0.15 & 0.0 & 0.01864 & 0.2774 \\ 
0.0 & 0.0 & 0.013446 & 0.2691 \\ 
-0.15 & 0.0 & 0.00965 & 0.2631 \\ 
-0.3 &+0.1& 0.00796 & 0.2577 \\ 
-0.5 &+0.2& 0.00593 & 0.2533 \\ 
-1.0 &+0.3& 0.00224 & 0.2493 \\ 
\hline
\end{tabular}
\label{tab:abund}
\end{center}
\end{table}

\subsection{Nuclear reaction rates and network}
The nuclear energy production is computed with a 
reaction network including 185 nuclear reactions involving 54 stable and unstable species from $^{1}$H to $^{37}$Cl. We essentially use the same rates as in \cite{Lagarde2012} except for nuclei with mass number $A < 16$, for which we adopt the updated rates from the NACRE II compilation \citep{Xu2013b}. The numerical tables used in the code are generated using the NetGen web interface\footnote{\url{http://www.astro.ulb.ac.be/Netgen/form.html}} \citep{Xu2013a}. The screening factors are calculated with the formalism of \citet{Mitler77} for weak and intermediate screening conditions and  of \citet{Graboske73} for strong screening conditions. 

\subsection{Treatment of the atmosphere}
\label{Sect:Atm}
A special attention was given to the treatment of the stellar atmosphere. In the STAREVOL code, the stellar structure equations are solved in one shot from the center to the surface: there is no decoupling between the interior and the atmosphere as it is done in some stellar evolution codes. The surface boundary conditions are treated using the so-called \emph{Hopf function}, $q(\tau)$ that provides at a given optical depth $\tau$ a correction to the grey approximation (see \citealt{Hopf30,Morel1994}) :
\begin{equation}
\frac{4}{3}\left(\frac{T(\tau)}{T_{\mathrm{eff}}}\right)^4 = q(\tau) + \tau,
\label{Eq:qtau}
\end{equation}
where $T_{\rm eff}$ is the temperature of the equivalent black body and $T(\tau)$ the temperature profile. 
In the previous PMS grid, \citet{Siess2000} used analytical $q(\tau)$ expressions derived from tailored Kurucz and MARCS model atmospheres. 

In the present study, the functions $q(\tau)$ are calculated from the values of $T(\tau)$ and $T_{\rm eff}$ given by the PHOENIX atmosphere models \citep{Allard2012}. We selected these models\footnote{\url{http://perso.ens-lyon.fr/france.allard/}} because of their wide coverage in $2600~K\leq T_{\rm eff}\leq 70000~K$, $0\leq\log g$(cm.s$^{-2}$) $\leq$ +5.0 and $-4.0\leq$ [M/H] $\leq +0.5$) and also because they adopt the same solar mixture \citep{AsplundGrevesse2009} and a mixing-length parameter value $\alpha_c = 2.0$ very close to ours of $1.973$.
The connection between the atmosphere and the interior can be done at a specific temperature \citep[e.g.][]{FeidenChaboyer2012}, or at a given optical depth $\tau_{ph}$ \citep[e.g.][]{Tognelli2011,Chenetal2014,BHAC15}.
As shown in \cite{Montalban2004}, the results of the calculations does not depend sensitively on the location of the  matching point provided it remains in a region where $10 < \tau_{ph} < 100$. However, in STAREVOL, we consider that each point between $\tau=30$ and $\tau=100$ is a matching point. So, given the metallicity and surface gravity of the model, we search for the model atmosphere effective temperature that give the stellar structure temperature at the matching point's  optical depth. We then calculate an average effective temperature from the previously computed values of $T_{\rm eff}$. Finally, we interpolate in the grid of model atmosphere the new temperature profile corresponding to the mean $\langle T_{\rm eff} \rangle$  and use Eq.~\ref{Eq:qtau} to obtain the expression of $q(\tau)$. 
This model atmosphere, which is calculated at each iteration during the convergence process, also provides the other outer boundary condition on the density. In our calculations, we set the numerical surface to be at an optical depth  $\tau = 0.015$.

\begin{figure}
\includegraphics[width=0.45\textwidth]{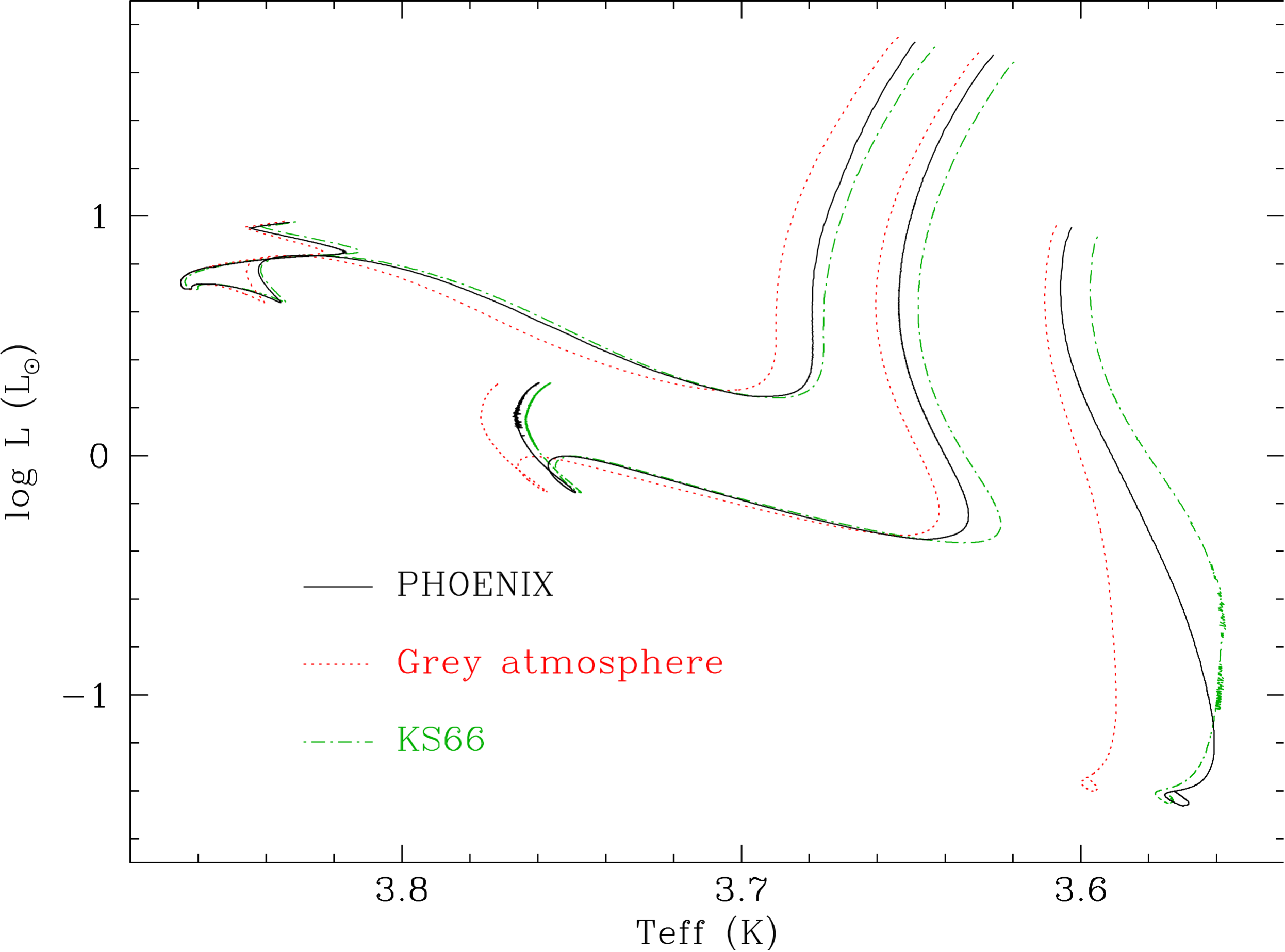}
\caption{Evolutionary tracks in the Hertzsprung-Russell diagram for standard models of 0.5, 1.0 and 1.5\Ms at $Z_\odot$ from bottom to top using three different boundary conditions as indicated on the plot. KS66 refers to the \cite{KrishnaSwamy} prescription.}
\label{Fig:hrd_atm}
\end{figure}

The effect of changing the surface boundary condition is illustrated in Fig.~\ref{Fig:hrd_atm}. 
The 1.5\Ms model is almost unaffected by the use of a realistic atmosphere model, except on the Hayashi track, where the tracks are 100~K cooler than in calculations using a grey atmosphere boundary condition. This difference is also present on the 1.0\Ms Hayashi track and remains along the main sequence. As expected, the colder the surface and hence the lower the stellar mass, the stronger the impact of the atmosphere on the structure and hence on the location of the tracks in the HRD. For the 0.5\Ms case, the difference between a model using a grey atmosphere and one using a PHOENIX atmosphere can exceed  300~K during the PMS and  200~K on the MS. We note that the models using \cite{KrishnaSwamy} prescription fit quite well the evolution with PHOENIX atmosphere models at solar metallicity even in the low-mass regime. 

\subsection{Mixing length parameter for convection}
\label{mlt}
The use of new boundary conditions and new input physics requires a new calibration of the mixing-length parameter (and initial helium content, see \S~\ref{Sect:initabund}) to reproduce the solar radius and luminosity at the current age of the Sun. We calibrated the luminosity and radius of a non-rotating 1.0\Ms model, neglecting mass-loss, to a relative precision of $10^{-4}$. 
We obtain a mixing-length parameter value $\alpha_c=1.973$ with a helium content $Y=0.269$ for a metallicity $Z=0.0134$ corresponding to the \citet{AsplundGrevesse2009} mixture. We keep the same value of $\alpha_c$ for all the models of the grid.
In all our models -- standard and rotating -- overshooting is not considered.

\subsection{Mass-loss prescriptions}
We compute the mass-loss rate all along the evolution with two different prescriptions depending on whether it is a rotating model or not. In the non-rotating case, the mass-loss rate follows \citet{Reimers1975} (with $\eta_R=0.5$) while in the rotating case we use the mass-loss rate given by \citet{CS2011}. The latter takes into account the effects of rotation on the stellar activity and thus depends implicitly on the stellar spin. A metallicity scaling is applied to the mass loss rate following \cite{Mokiem2007} : 
\begin{equation}
\dot{M} = \dot{M} \times \left(\frac{Z}{Z_\odot}\right)^{0.8}\ .
\end{equation}

\subsection{Angular momentum evolution}
\subsubsection{Initial spin velocity}
\label{sec:initrot}

{The initial models are build from polytropes and are all fully convective. We assume an initial solid body rotation and consider 3 different initial rotation period (P$_\mathrm{rot,init} = 1.6$, 4.5 and 9 days) rate for which we associate a disc locking timescale. The values of $\tau_\mathrm{DC}$ and P$_\mathrm{rot,init}$ (see Table~\ref{tab:rot}) are based on the study of \citet{GB15} and are chosen to reproduce the  spread in rotation periods observed in young open clusters \citep[see][]{GB13,GB15,Amard2016,SadeghiA2017}. Let us mention that, in order to prevent the most massive stars (1.2-1.5\Ms) from exceeding their critical velocity, we increased the initial rotation period from 1.6 to 2.3 days and the disc-coupling duration from 2.5 to 4~Myr for the fast rotating models.}

{Our simplified treatment of disc-coupling ($\tau_\mathrm{DC}$ independent of mass and metallicity for a given P$_\mathrm{rot,init}$) implies that the rotation period remains constant during the first few million years of evolution. 
Our initial models can have very large radii and for the initially fast rotators, the spin acceleration following the star-disc unlocking, may bring the star to break-up velocities.
To avoid this non-physical situation, the models are evolved as non-rotating for the first five hundred thousand years  (which sets the time zero of the evolution of our models) and only after this time, is rotation taken into account. Such a precaution is not necessary for the median and slow rotators.}

Finally, as in \citet{Amard2016}, the effects of rotation on the structure are treated following the formalism of \citet{ES76}.

\subsubsection{Internal transport of angular momentum}
\label{Sect:IntTransp}
We describe the transport of AM in the stellar interior following the formalism of \citet{Zahn1992} as updated by \citet{MaederZahn1998} and \citet{MathisZahn2004}. The transport of AM happens on a secular timescale in the radiative regions of the star \citep[see e.g.][]{DecressinMathis2009}. As the star evolves, differential rotation develops in the radiative region that contribute to the transport of AM between the core and the envelope. We assume that the convective regions rotate as a solid body.

Based on the recent work on the anisotropy of turbulence in stellar radiative regions by \cite{Mathis2018}, we modified the set of prescriptions for the turbulent diffusion coefficients compared to what was used in \citet{Amard2016}. The horizontal turbulence ($\nu_{\rm h}$) is now the sum of two terms, one ($\nu_{\rm h,v}$) corresponding to the component created by the vertical shear, and one ($\nu_{\rm h,h}$) corresponding to the shear that develops along the isobar. 
$\nu_{\rm h,v}$ is set to 0 when the vertical shear is not important enough to fulfill the Reynold's criterion (i.e., $\nu_{\rm v} \geq \nu_{\rm m} Re_c$ where $Re_c = 7$; Prat, V., Private Communication). For consistency with the expression of the horizontal turbulence, we use \citet{Zahn1992} prescription for the vertical shear-induced turbulence $\nu_{\rm v}$.
These prescriptions do not require any parameter fine-tuning over the  mass, rotation, and chemical composition ranges covered by our grid. 

We recall the advection-diffusion equation for the transport of AM as given in \citet{Zahn1992} and \citet{MathisZahn2004}
\begin{equation}
\rho \frac{{\rm d}}{{\rm d}t}\left(r^2\Omega\right)= \frac{1}{5r^2}\ddr{} \left(\rho r^4 \Omega U_r\right) + \frac{1}{r} \ddr{} \left(r^4\rho \nu_v \ddr{\Omega}\right),
\label{eq:general}
\end{equation}
where $\rho$, $ r$, $\nu_v$ and $U_r$ are the density, radius, vertical component of the turbulent viscosity, and the meridional circulation velocity on a given isobar, respectively. 
By integrating this equation at a given radius $r$ we obtain a flux equation,
\begin{equation}
F_{\rm tot} = F_{S}(r) + F_{MC}(r)
\label{eq:fluxeq}
,\end{equation}
with
\begin{equation}
F_{S}(r) = \frac{\mathrm{d} J_{S}}{\mathrm{d} t} \bigg|_{r} = -\rho r^4 \nu_v \ddr{\Omega}\bigg|_{r}
\label{eq:Fshear}
\end{equation}
the flux carried by shear-induced turbulence from the radiative zone to the convective envelope (CE), and
\begin{equation}
F_{MC}(r) = \frac{\mathrm{d} J_{MC}}{\mathrm{d} t} \bigg|_{r}= -\frac{1}{5} \rho r^4 \Omega U_{r}
\label{eq:Fcirc}
\end{equation}
the flux carried by meridional circulation. A detailed derivation of the AM fluxes is given in \cite{DecressinMathis2009} as part of a set of tools for assessing the relative importance of the processes involved in AM transport in stellar radiative interiors.

Nevertheless, we would like to put a word of caution. The close-to-breakup stars and their internal transport are not expected to be rigorously modeled with our formalism because some assumptions are not fulfilled anymore. More careful work would imply the use of at least two dimensions simulations that are not available as of today for the considered evolutionary timescales \citep{EspinosaLaraRieutord2007,Hypolite2018}. 

\subsubsection{Extraction of angular momentum}
\label{sec:extAM}
From the birthline to the TAMS, the AM content decreases by two orders of magnitude as the result of two main processes.

\subsubsection*{\small Disc coupling during early evolution}

Young stars are spun up by contraction and  acccretion of AM through their circumstellar disk. But they are also braked by the development of accretion-induced winds \citep[][]{MattPudritz2005,MattPudritz2008,ZanniFerreira2009}, magnetospheric ejections \citep[see \emph{e.g.}][]{Shu1994,ZanniFerreira2013} or by the so-called disc-locking process \citep[see \emph{e.g.}][]{GhoshLamb1979}. Observations  \citep{Rebull2004,GB13} indicate that  the interaction is very efficient and results in  an almost constant stellar angular velocity  during the disc lifetime. With this assumption of strong coupling, angular momentum evolution models \cite[e.g.][]{BouvierPPVI} are able to reproduce the overall spread in surface velocities provided the disc lifetime is not unique.

As reported in several studies \citep{KennedyKenyon2009,WilliamsCieza2011,VasconcelosBouvier2017}, the duration of the disc-locking phase is likely dependent on the stellar mass, initial rotation period \citep{GB13}, or stellar chemical composition \citep{Ghezzi2018}. In the absence of a clearer view, we use a unique disc coupling timescale ($\tau_\mathrm{DC}$ in table~\ref{tab:rot}) for all stars that depend only on the initial rotational period.

Additional AM loss due to magnetic wind braking is also considered  all along the evolution as described in the coming section.

\subsubsection*{\small Extraction of angular momentum by stellar winds}
Low-mass stars with an external convective envelope sustain a dynamo-generated magnetic field, and thus undergo efficient magnetic braking during their evolution through their magnetized wind \citep[\emph{e.g.}][]{Schatzman1962}. 
While the prescription by \citet{Kawaler88} has been extensively used to account for AM loss, recent theoretical studies \citep{RM12,Matt2012,Mattetal2015,VSP2013,Vansaders2016,Reville2015a,Garraffo2018} provide a variety of so-called ``braking law'' that take into account various physical ingredients and are calibrated on different observational samples.
In our models, the torque applied at the surface of the star is computed following \citet{Mattetal2015} formulation and writes 
\begin{equation}
\frac{\mathrm{d}J}{\mathrm{d}t} = -\mathcal{T}_0 \left(\frac{\tau_{cz}}{\tau_{cz\odot}}\right)^{p} \left(\frac{\Omega_\star}{\Omega_\odot}\right)^{p+1}  \rightarrow {\rm unsaturated},
\label{eq:torque_unsat}
\end{equation}
\begin{equation}
\frac{\mathrm{d}J}{\mathrm{d}t} = -\mathcal{T}_0 \chi^p \left(\frac{\Omega_\star}{\Omega_\odot}\right)  \rightarrow  {\rm saturated},
\label{eq:torque_sat}
\end{equation}
with 
\begin{equation}
\mathcal{T}_0 = K \left(\frac{R_\star}{R_\odot}\right)^{3.1} \left(\frac{M_\star}{M_\odot}\right)^{0.5}\gamma^{2m}  ,
\label{Eq:torque0}
\end{equation}
where $\gamma = \sqrt{1+(u/0.072)^2}$ comes from Eq. $(8)$ of \cite{Matt2012}, and $u$ is the ratio of the surface velocity to the brake-up velocity. The calibration constant $K$ is expected to be close to the solar wind torque derived from spin models \citep{Finley2018} and $\Omega_\star$ is the surface angular velocity with $J$ the stellar angular momentum. $R_\star$ and $M_\star$ denote the radius and stellar mass and the symbol $\odot$ indicates the solar value.
This formalism depends on the Rossby number in the convective envelope \citep[][]{Nandy2004,Jouve2010}, defined as 
\begin{equation}
Ro = 1/(\tau_{cz}\Omega_\star),
\label{Eq:Rossby}
\end{equation}
with  $\tau_{cz}$ the turnover timescale estimated at 0.5 pressure scale height above the base of the convective envelope. In our model, the magnetic field  saturates when $Ro < Ro_{\rm sat}$ and this saturation value is determined by imposing that our 1\Ms roughly reproduces the dispersion in rotation periods in the $\alpha$-Per ($\approx$ 85 Myr) and M35 ($\approx$ 150 Myr) open clusters. This requires $\chi = \frac{Ro_\odot}{Ro_{\rm sat}} = 14$ with $Ro_\odot \sim 2$, and thus a saturation Rossby number $Ro_{\rm sat}=0.14$ very close to 0.13$\pm$0.02 as observationnaly derived by \citet{Wright2011}. 

One may wonder if it is realistic to derive convective velocities from a formalism as simple as the mixing-length theory. Multi-dimensional simulations of convection \citep{Hanasoge2012,Viallet2013,trampedach2014b} have been showing that the mixing-length theory provides good estimates for convective velocities. This is particularly true close to the bottom of the convective envelope, where we probe the convective turnover timescales, thus making our derived Rossby numbers more reliable.

In \citet{Amard2016}, we used the parametric relation between $Ro$ and the effective temperature as suggested by \citet{CS2011}.
We refer the reader to \citet{Charbonnel2017} for a description of the variations of this quantity within the stellar convective envelope along the evolution and as a function of stellar mass and metallicity.

\subsubsection*{\small Torque calibration on observational constraints}
With the updated physics, the constant $K$ (Eq.~\ref{Eq:torque0}) had to be re-calibrated to reproduce the Sun's rotation rate. We also calibrate by eye the value of $p$ (Eqs~\ref{eq:torque_unsat},\ref{eq:torque_sat}) to match the observed velocity dispersion in the Pleiades and Praesepe clusters for the 1.0\Ms and 0.5\Ms models.
The adopted values of the parameters $K,p,m$ and $\chi$ are given in Table~\ref{tab:brake} and are kept constant over the entire mass and metallicity range, independently of the initial rotational velocity. 

\begin{table}[h]
\begin{center}
\caption{Parameters used for \citet{Mattetal2015} prescription.}
\begin{tabular}{ c  c  c}
\hline
\hline
{\sf  Parameter} & \cite{Amard2016} & Present work\\
\hline
$K$ & 5 10$^{31}$ & 7 10$^{30}$ \\ 
$m$ & 0.22  & 0.22\\ 
$p$ & 1.7 & 2.1\\ 
$\chi$ & 10 & 14\\ 
\hline
\end{tabular}
\label{tab:brake}
\end{center}
\end{table}

\subsection{Transport of chemicals}

In the rotating low-mass stars, rotation-induced mixing is expected to erase the effect of atomic diffusion (see Deal et al. in prep) because of the presence of a relatively thick surface convection zone. In these stars, the efficient braking of the surface by the magnetized stellar winds generates  a strong shear below the convective envelope, responsible for an efficient mixing of the chemical species.
For stars with a very shallow convective envelope, \ie with a larger mass and/or lower metallicity, the differential rotation will be reduced and radiative levitation will become the main agent of chemical mixing \citep[\eg][]{Richard2002}.
Since we do not account for neither gravitational settling nor radiative levitation, the surface composition of our models with M > 0.8 \Ms is not expected to be realistic 
(see \S~\ref{sect:Lithium}). A full and consistent treatment of chemical transport including rotational induced mixing and atomic diffusion is one of our priority for a forthcoming study.

The vertical transport of nuclides in the radiative regions results from the combined action of meridional circulation and turbulent shear whose formulation follows\cite{chaboyerzahn92}.
For a chemical species $i$, the concentration $c_i$ obeys 
\begin{equation}
    \frac{\mathrm{d}c_i}{\mathrm{d}t} = \dot{c}_i + \frac{1}{\rho r^2} \frac{\partial}{\partial r}\left(\rho r^2 D_\textrm{tot} \frac{\partial c_i}{\partial r}\right),
\end{equation}
where $D_\mathrm{tot} = D_\mathrm{eff} + D_\mathrm{v}$ is the total diffusion coefficient and $D_\mathrm{eff}$ is given by 
\begin{equation}
    D_\mathrm{eff} = \frac{|rU(r)|^2}{30 D_\mathrm{h}},
\end{equation}
where $D_\mathrm{v}$ and $D_\mathrm{h}$ are the vertical and horizontal turbulent diffusion coefficient, respectively. 
The term $\dot{c}_i$ accounts for the evolution of the concentration of chemical species $i$ due to nuclear burning.


\section{Description of online electronic tables }
\label{sec:web}

\begin{center}
\begin{table}
\caption{Grid parameters.}
\begin{tabular}{ c    c        c       c       c     }
\hline \hline
Mass (\Ms) & \multicolumn{4}{c}{0.2 - 1.5 (0.1 steps)$^\star$} \\
\hline
([Fe/H]) & \multicolumn{4}{c}{-1, -0.5, -0.3, -0.15, 0.0, +0.15, +0.3} \\
\hline
& \it{fast} & \it{median} & \it{slow} & \it{standard} \\
P$_{\rm rot,init}$ (days) & 1.6  (2.3)$^\dagger$ & 4.5 & 9.0 & - \\
$\tau_{\rm DC}$ (Myr)& 2.5 (4)$^\dagger$ & 5 & 5 & - \\
\hline 
\end{tabular}
\tablefoot{$^\dagger$ Values used for the 1.2 to 1.5\Ms models.\\ $^\star$ a step of 0.05 is used in the mass interval [0.7\Ms; 1.3\Ms].}
\label{tab:rot}
\end{table}
\end{center}

\begin{figure}
\includegraphics[width=0.48\textwidth]{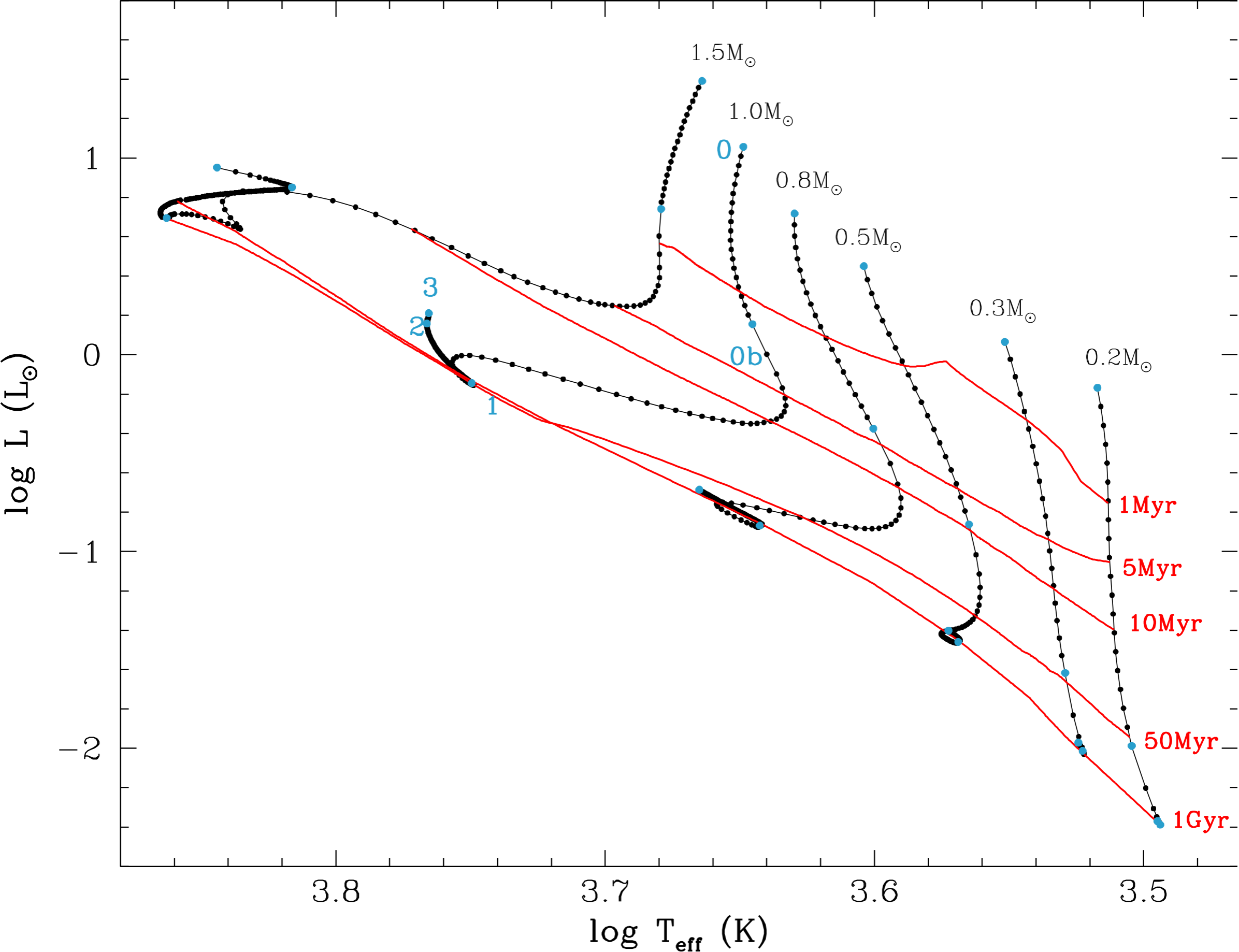}
\caption{Evolution track in the Hertzsprung-Russell diagram of six standard models of 0.2, 0.3, 0.5, 0.7, 1.0 and 1.5\Ms at solar metallicity. Isochrones corresponding to 1, 5, 10, 50 Myr and 1 Gyr are also represented in red. In blue are shown the evolutionary points described in the text.}
\label{fig:Hr_ptsgrilles}
\end{figure}

Our models have been integrated in the \textsc{Syclist} toolbox \citep{Georgy2014b}.
The published tables include 100 new points to describe the PMS evolution and two new key points to the list described in \citet{Ekstrom2012}. The first key point defines the beginning of the pre-main-sequence phase (the point where the star is $10^5\,\text{yr}$ old). The second one indicates where the radiative core appears. We labelled these key points ``0'' and ``0b'', so that key point labelled ``1'' (defined as the ZAMS where the star has burnt $0.003$ hydrogen in mass fraction) is the same as in \citet{Ekstrom2012}.\\

For each model, we have selected a total of 290 points to allow a good description of the full tracks. We recall here the different key evolutionary steps (see Fig.~\ref{fig:Hr_ptsgrilles}): 
\begin{itemize}
\item[0 ] beginning of the pre-main sequence (1-20) ;
\item[0b] appearance of the radiative core if relevant (21-100 pts) ;
\item[1 ] ZAMS (101-185 pts) ;
\item[2 ] turning point with the lowest $T_\mathrm{eff}$ on the main sequence (186-210) ;
\item[3 ] Main sequence turn-off ;
\end{itemize}

Point 0 exists for all models, as well as point 1. Point 0b is not defined for very low mass stars and in this case,  we set it to same pre-main-sequence time fraction as in the lowest mass model (of same metallicity and initial velocity) where it appears. There are 18 points between key point 0 and keypoint 0b, regularly spaced in terms of $log(L)$, so that key point 0b is the 20th point in the table. We then put 80 other points between key point 0b and keypoint 1, so that the ZAMS (which is the first point in the tables from \citet{Ekstrom2012}) is now the 101st point in the table. The points between key point 0b and key point 1 are equally spaced in time. For stars that do not reach the turn-off by 15 Gyr, we set point 3 to the last computed point. And point 2 is set as the last computed point for stars that do not reach the main sequence within 15 Gyr.
For each model, the quantities given in Table \ref{tab:gridtable} in the annexe, can be retrieved from the Geneva webpage\footnote{\url{https://www.unige.ch/sciences/astro/evolution/en/database/}}. 
We also provide the conversion of each track in two photometric system. The conversion in GAIA colours come from \citet{Evans2018} and the ones in the Johnson-Cousin system follow \citet{WortheyLee2011}.

\subsection*{Asteroseismic quantities}
All our stars have a convective envelope during the main sequence that is expected to generate solar-like oscillations. Following \citet{Lagarde2012}, we provide different global asteroseismic parameters listed in  Table \ref{tab:gridtable} that are computed from the structural properties of the models at each timestep. 
They include a number of scaling relation, the large separation from scaling relation 
\begin{equation}
    \Delta\nu_\mathrm{scale} = \Delta\nu_\odot \left(\frac{M}{M_\odot}\right)^{0.5} \left(\frac{R}{R_\odot}\right)^{-1.5},
\end{equation}
the frequency with the maximum amplitude 
\begin{equation}
    \nu_\mathrm{max} = \nu_{\mathrm{max},\odot} \left(\frac{M}{M_\odot}\right) \left(\frac{R}{R_\odot}\right)^{-2}\left(\frac{T_\mathrm{eff}}{T_\mathrm{eff}}\right)^{-0.5},
\end{equation}
the maximum amplitude  
\begin{equation}
    A_\mathrm{max} = A_{\mathrm{max},\odot} \left(\frac{L}{L_\odot}\right)^{0.838} \left(\frac{M}{M_\odot}\right)^{-1.32} \left(\frac{T_\mathrm{eff}}{T_\mathrm{eff}}\right)^{-2},
\end{equation}
with $\Delta\nu_\odot=134.9\mu$Hz, $\nu_{\mathrm{max},\odot}=3150\mu$Hz, and $A_{\mathrm{max},\odot} = 2.5$ppm the solar values.

Some asymptotic asteroseismic quantities are also provided:
the asymptotic large separation 
\begin{equation}
    \Delta\nu_\mathrm{asymp} = \left(2\int^R_0 \frac{\mathrm{d}r}{c_s}\right)^{-1},
\end{equation}
with $R$ the stellar radius and $c_s$ the sound speed, 
the total acoustic radius ($T$), 
\begin{equation}
    T = \int^R_0 \frac{\mathrm{d}r}{c_s} = \frac{1}{2 \Delta\nu_\mathrm{asymp}},
\end{equation}
the acoustic radii at the base of the CE ($t_\textrm{BCE}$) and at the location of the helium second-ionisation region ($t_\textrm{He}$). 
\begin{eqnarray}
    t_\mathrm{BCE} = \int^{r_\mathrm{BCE}}_0 \frac{\mathrm{d}r}{c_s}, && t_\mathrm{He} = \int^{r_\mathrm{He}}_0 \frac{\mathrm{d}r}{c_s},
\end{eqnarray}
with $r_\mathrm{BCE}$ and $r_\mathrm{He}$ the stellar radius at the base of the CE and of the helium second-ionisation region, respectively.
Finally, the asymptotic period spacing of g-mode defined as 
\begin{equation}
    \Delta\Pi (l=1) = \sqrt{2} \pi^2 \left(\int^{r_2}_{r_1} N \frac{\mathrm{d}r}{r}\right)^{-1}.
\end{equation}
where $r_1$ and $r_2$ are the radii that define the cavity where the g-modes are trapped and $N$ is the Brunt-V\"ais\"al\"a frequency. 
For more details, see the original article.

\subsection*{Isochrones}
We also provide the possibility to compute isochrones in different spectral bands with different filters.
These isochrones are computed using the \textsc{Syclist} tool and the reader is referred to \cite{Georgy2014b} for corresponding details. 

\section{Comparison with other grids at solar metallicity}
\label{sec:comp}

The important updates in the physics of the STAREVOL code since \citet{Siess2000} (see \citealt{Lagarde2012} and \S~\ref{sec:code}) justifies the computation of a new set of grids. Moreover, over the past few years several research groups have published  PMS evolutionary models. A comparison is therefore timely and will allow to assess the uncertainties in terms of HR diagram positions and ages associated with the use of different codes and input physics.

In Table~\ref{tab:parameter} we compile the main physical assumptions used in the computation of publicly available stellar evolutionary tracks. These models are standard, \textit{i.e.}, non-rotating and cover our grid mass range. 
In this table, the chemical mixture (column 2) refers to the adopted solar metallicity ($Z$).
We also provide the initial helium mass fraction $Y$ and the mixing length parameter $\alpha_{MLT}$ (columns 2 and 3 respectively). The solar symbol $\odot$ in column 3 indicates the grids that use a calibration of their solar model (in terms of luminosity and radius at the age of the Sun) to determine $Y$ and $\alpha_{MLT}$.
In column 4 we indicate the set of model atmospheres used as external boundary condition and the optical depth where they are attached to the stellar interior.
The adopted equation of state (thereafter EOS) is given in column 5; the importance of its accuracy for PMS stars has been largely discussed in the literature \citep[\emph{e.g.}][]{Baraffe98,Siess2000}. 
In column 6 we recall the bibliographical sources for radiative opacities at high (first line) and low (second line) temperature. Most of the current evolution codes use OPAL radiative opacities tables for the interior computation where T > 8000K, but for T < 8000K, two main opacity tables are considered : \citet{AF94} and \citet{F05}.
Column 7 gives the source for the nuclear reaction rates while information about the use (or not) of core overshooting in the grid computation is given in column 8. The last two columns of the table give the age of the 1\Ms, $Z_\odot$ of each grid at the ZAMS and TAMS (see definition in the table notes) in Gyr, and the radius of this model at the ZAMS in units of \Rs. 

\begin{figure*}
\begin{center}
\includegraphics[width=0.33\textwidth]{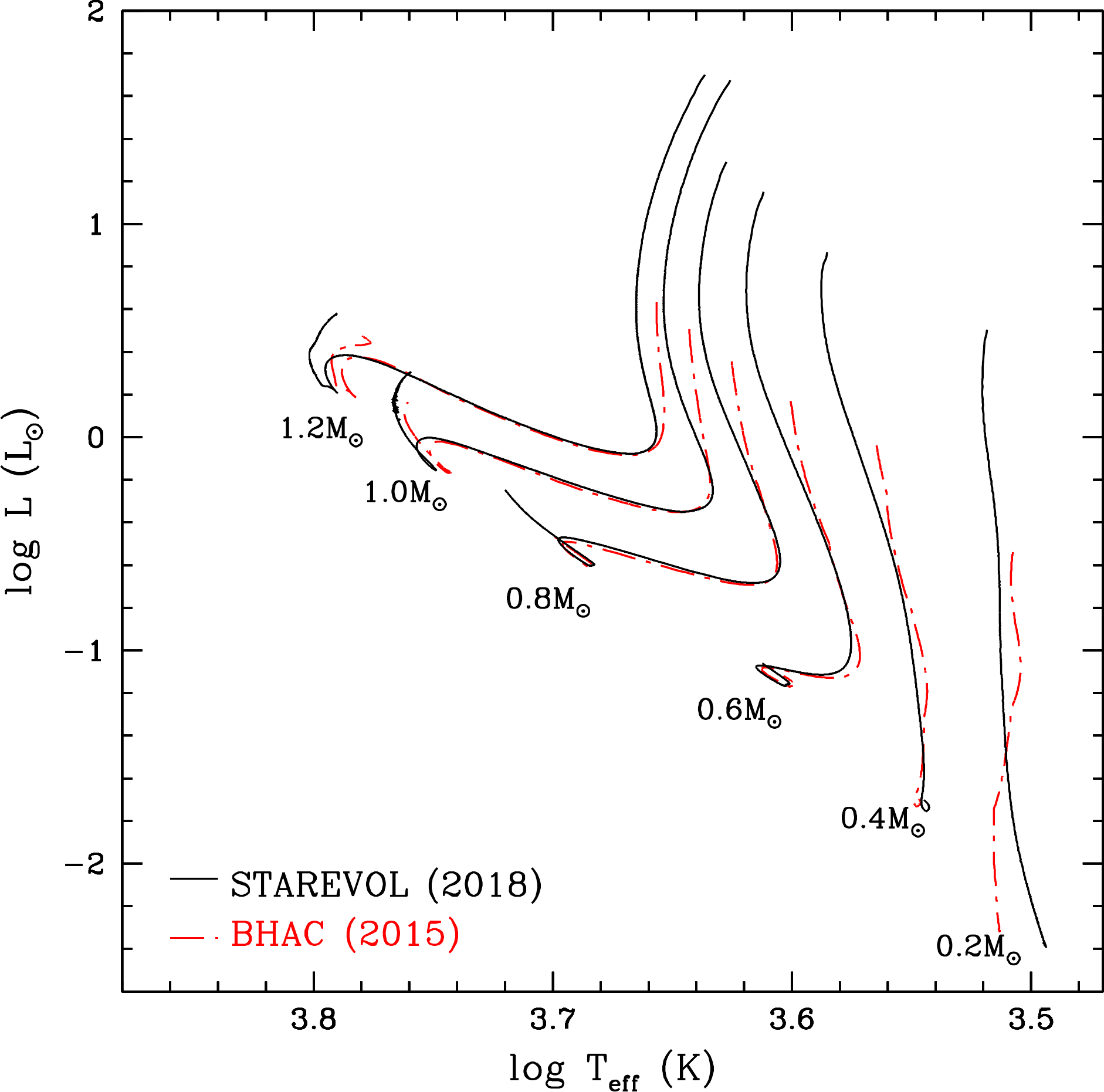}
\includegraphics[width=0.33\textwidth]{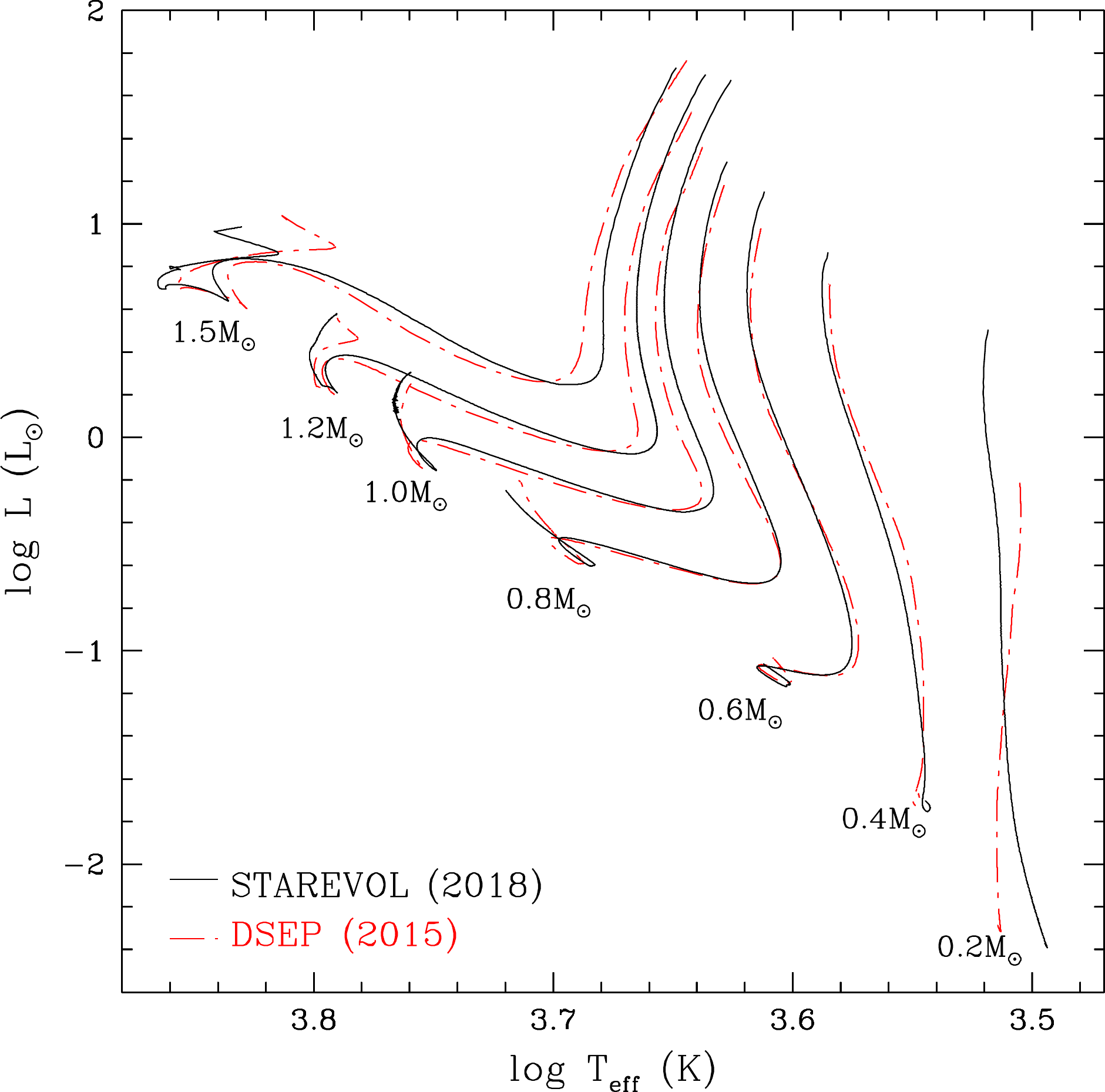}
\includegraphics[width=0.33\textwidth]{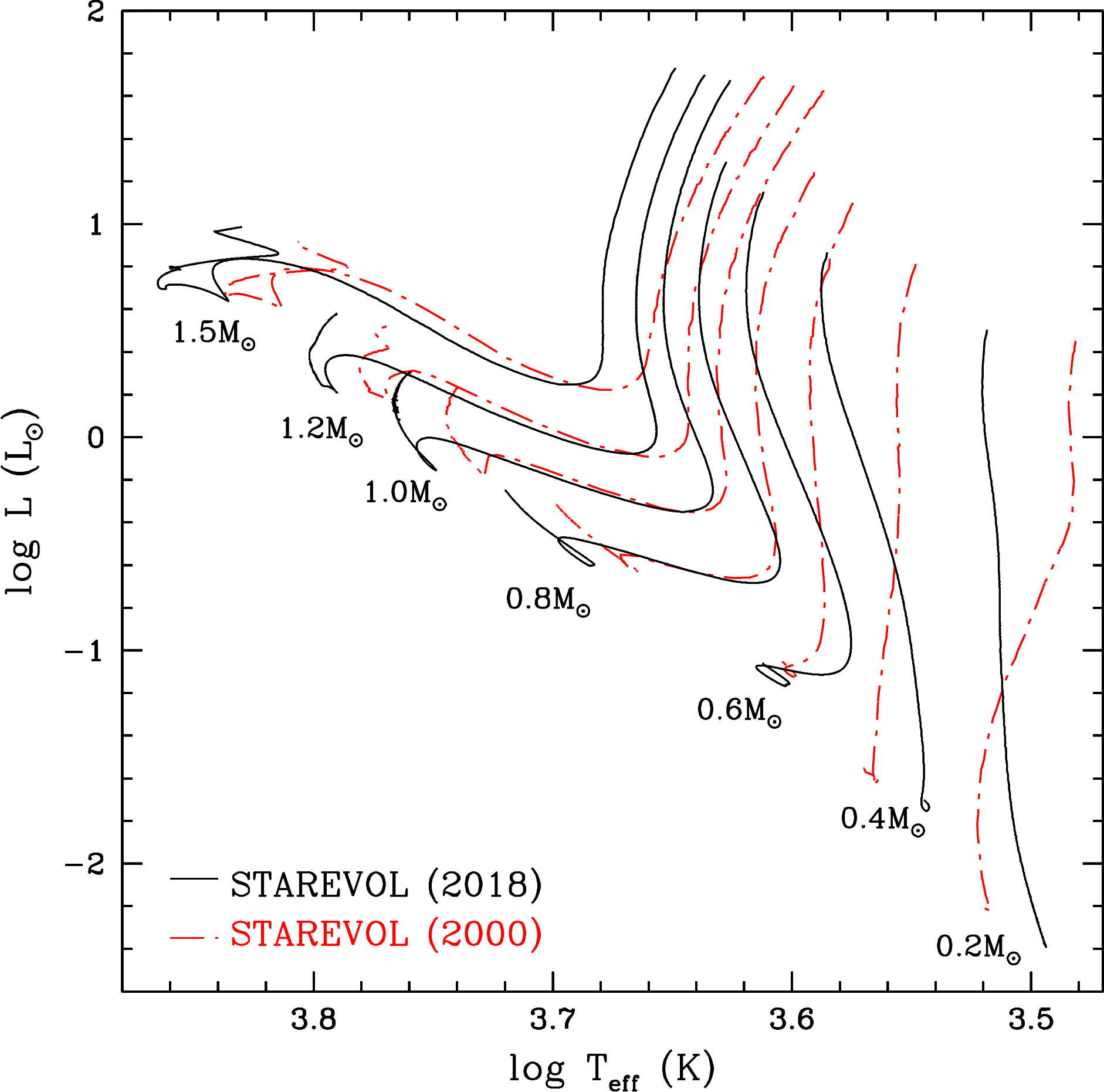}
\includegraphics[width=0.33\textwidth]{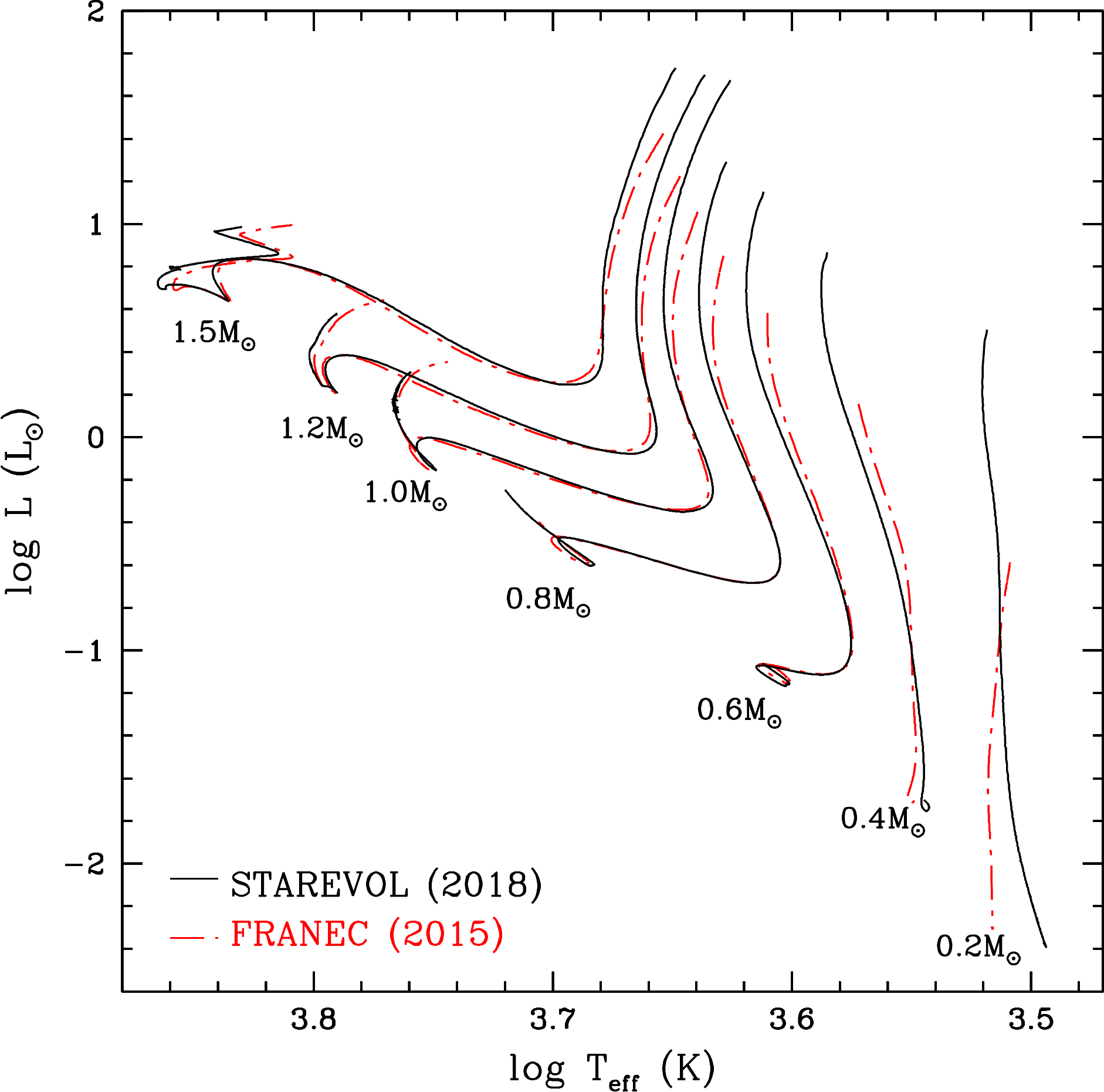}
\includegraphics[width=0.33\textwidth]{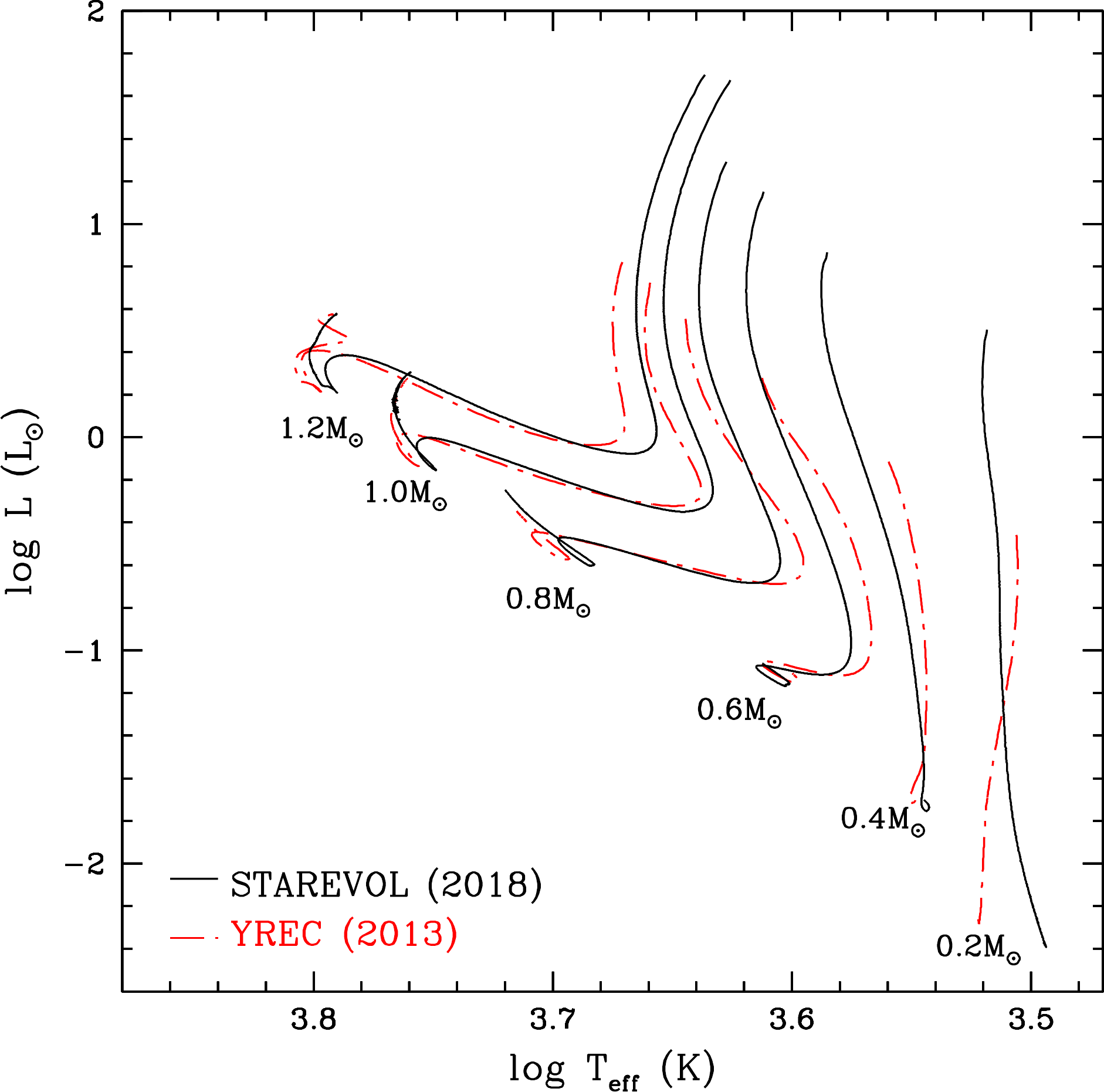}
\includegraphics[width=0.33\textwidth]{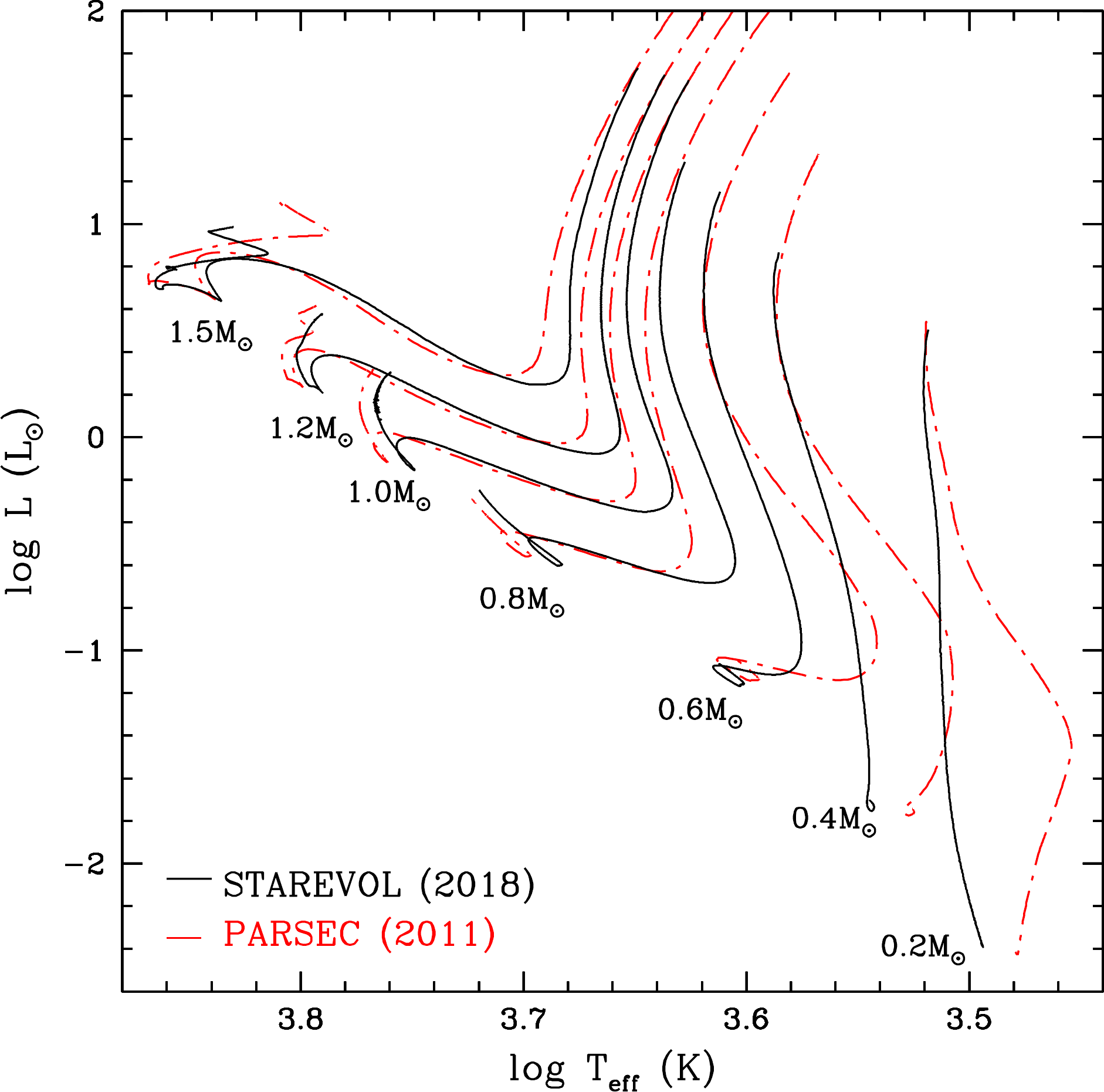}
\includegraphics[width=0.33\textwidth]{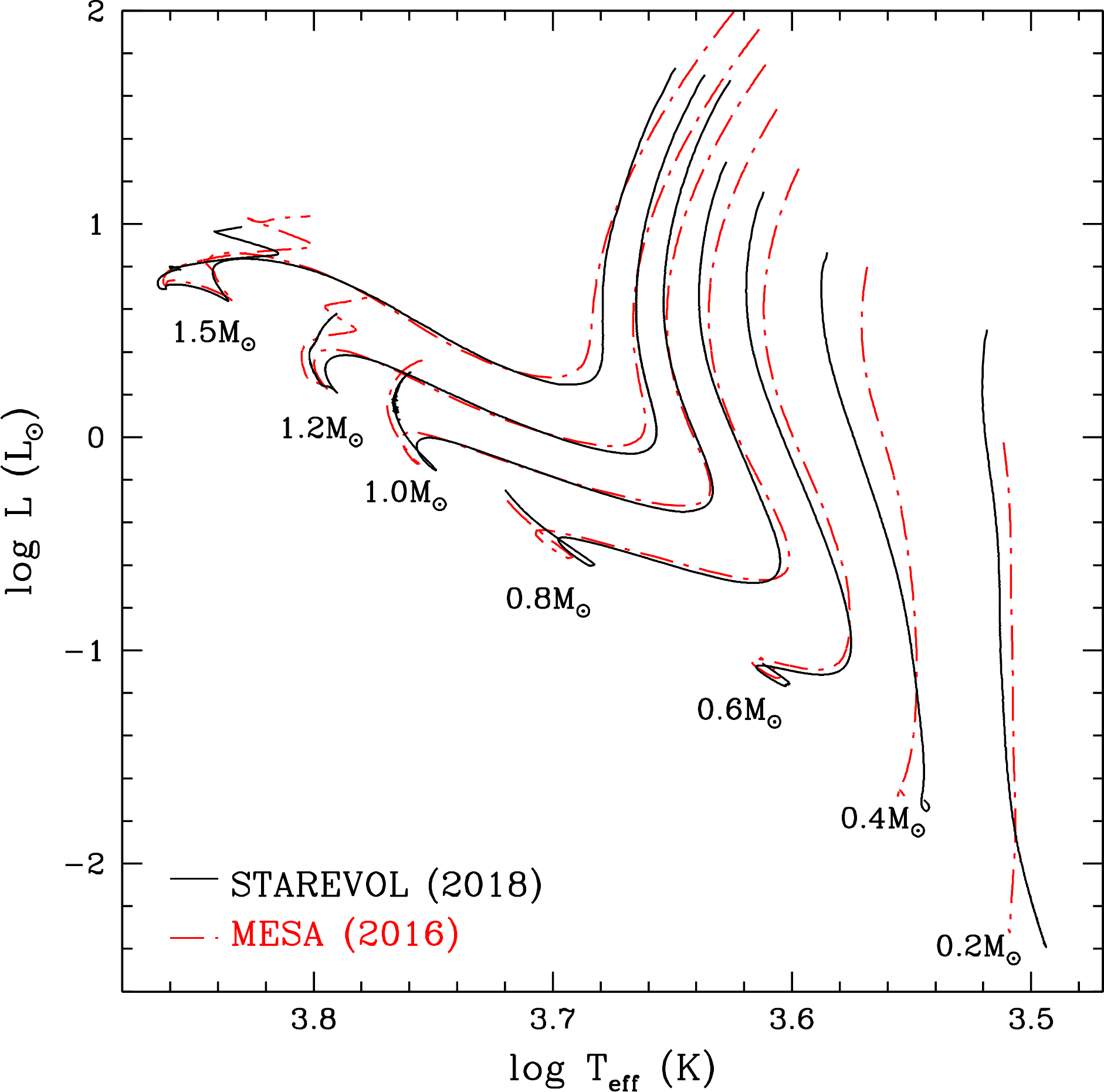}
\includegraphics[width=0.33\textwidth]{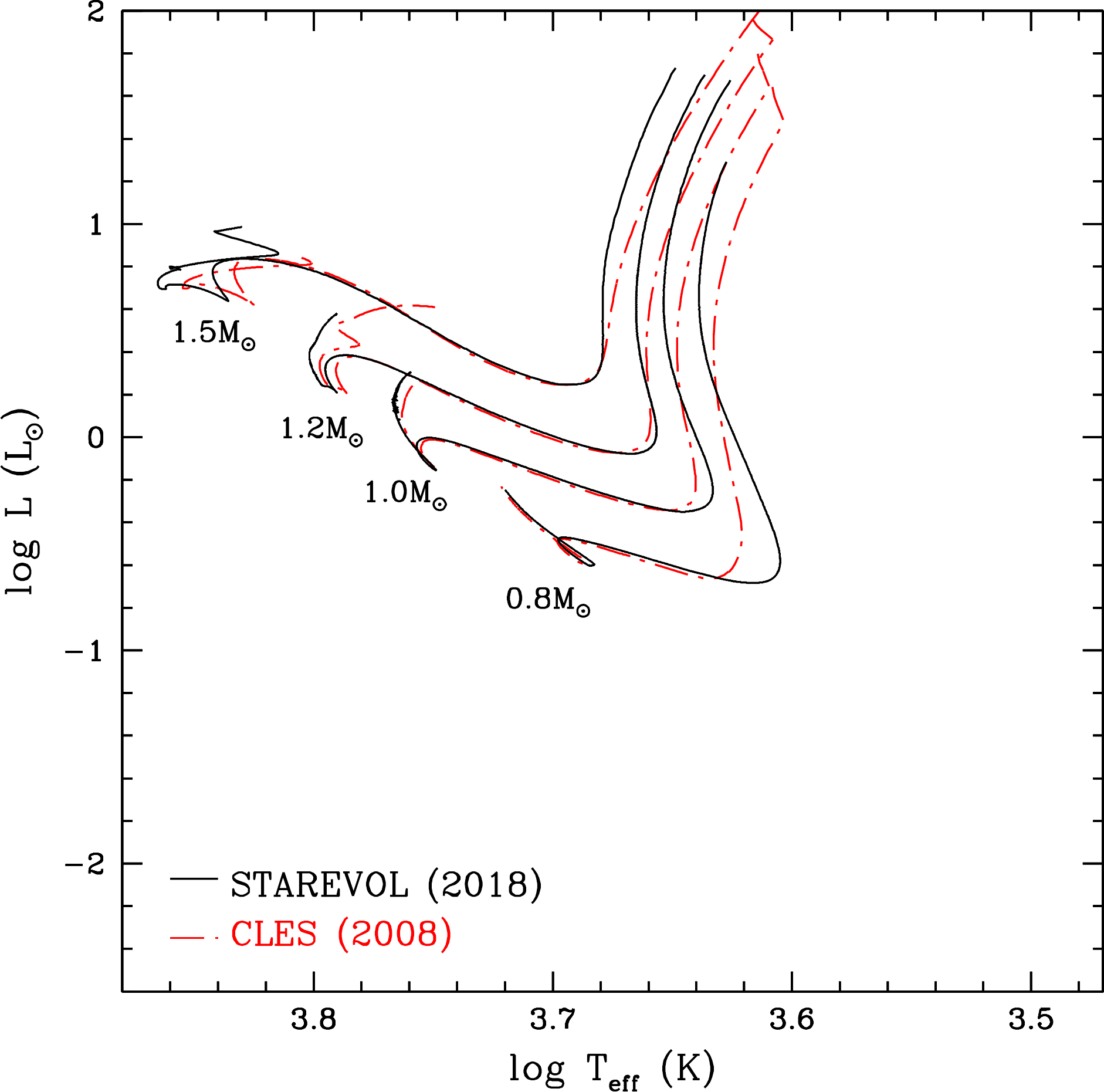}
\caption{Comparison of our standard solar metallicity models with other available grids as described in Table~\ref{tab:parameter} and indicated on each panel.}
\label{Fig:allhrd}
\end{center}
\end{figure*}

Below we compare in more details our grid of solar metallicity, standard, non-rotating models with the ones listed in Table~\ref{tab:parameter}. We find a good agreement with most of them (especially with FRANEC and MESA), as clearly visible from Fig.~\ref{Fig:allhrd} where we plot selected evolution tracks in the Hertzsprung-Russell diagram. There is a systematic offset between our 0.2\Ms model and others that we think is due to the new treatment of the upper part of the atmosphere that we use. 


\subsubsection*{STAREVOL : \cite{Siess2000}}
\citet{Siess2000} grid has been extensively used for the two decades. Due to the important improvements of the constitutive physics during that period of time, this oldest grid is also the one for which we find the largest discrepancies with our new models. This can be explained by the combined use of the \citet{GN93} solar reference abundances, with an associated metallicity $Z = 0.02$ much higher than our adopted value of $Z = 0.0134$, a smaller MLT parameter $\alpha$, and older atmosphere models boundary condition. For any given initial stellar mass, \cite{Siess2000} models are cooler than ours and the Henyey tracks are always shorter. This behaviour has already been discussed in \citet{Montalban2004} and according to them, is essentially due to an interplay between the mixing-length parameter, the chemical composition, and the peculiar atmosphere models. 

\subsubsection*{BHAC : \cite{BHAC15}}
The models by \citet[][hereafter BHAC]{BHAC15} are an updated version of \cite{Baraffe98} computed with an improved atmospheric treatment and the solar chemical mixture derived by \citet{Caffau11}. \cite{Baraffe98} were the first group to publish models that {\em self-consistently} couple the stellar interiors and state-of-the-art atmosphere models, therefore becoming a reference for low-mass stellar evolution models. Such an approach has since then been adopted by other groups, including ours. BHAC grid ensures the consistency of the convection treatment between the interior and the atmosphere, with a calibrated mixing length parameter. As shown in Fig.~\ref{Fig:allhrd} our evolution tracks are very close in the mass range 0.4 to 1.2\Ms. For the very low mass model 0.2\Ms, the treatment of molecular species becomes important and the models deviate as our EOS does not account for molecules heavier than H$_2$ contrary to that used by BHAC.

\subsubsection*{DSEP : \cite{Dotteretal2008}, \cite{Feiden2015}}
The Dartmouth Stellar Evolution Program (DSEP) contains suitable physics for PMS models computation. The surface boundary conditions are derived from PHOENIX model atmospheres \citep{Hauschildt99}. The computations by \cite{Feiden2015} include overshoot for the stars that are able to maintain a convective core (CC) during most of their lifetime: at solar metallicity, only models more massive than 1.1\Ms are concerned (see their Table 1). The overshoot parameter beyond the CC is assumed to vary with the stellar mass (0.05, 0.1, 0.2 and $0.2 H_{\rm p}$ are chosen for the 1.2, 1.3, 1.4 and 1.5\Ms models respectively). The extension of the CC increases the amount of hydrogen available for nuclear burning and so, the main sequence duration. The agreement between our models and the DSEP ones is very good for the low-mass star but below $M < 0.4$\Ms our tracks are cooler indicating a slighlty less compact structure. This discrepancy may be attributed to differences in the EOS as in these objects non ideal effects become important. For masses above 1.2\Ms the addition of overshooting in the DSEP models leads to a difference in the main sequence evolution, which lasts longer and extends further toward the red in their case compared to our models.

\subsubsection*{YREC : \cite{Spadaetal2011}}
In this comparison, we use the grid computed with the non-rotating configuration of the Yale Rotating Evolutionary Code (YREC) which includes a specific EOS for low-mass stars (see Table~\ref{tab:parameter}). The differences between our models are small for masses higher than 0.4\Ms. Below this limit, YREC changes its EOS to \cite{SCVH95}, which is the same as that used by \citet{BHAC15}. Thus, the difference between their models and ours in this mass range are comparable to the one that we have with BHAC.

\subsubsection*{FRANEC : \cite{Tognelli2011}, \cite{Valle2015}}
We compare our models with the ones of \cite{Tognelli2011} and \cite{Valle2015} who have updated the Frascati RAphson Newton Evolutionary Code (FRANEC) version developed in Pisa to account for new abundances and realistic atmospheric conditions as described in Table~\ref{tab:parameter}. Even though we use different mixing length parameter, treatment of atmosphere, these models are the ones in closest agreement with our calculations.

\begin{landscape}
\begin{table}
\begin{center}
\title{grid parameters}
\begin{tabular}{ c c c c c c c c c c }
\hline
\hline
&&&&&&&&&\\
Model & Chemical & $\alpha_{MLT}$ & Atmosphere & EOS & Opacities & Nuclear & Overshoot & 1\Ms ZAMS & Radius at \\
 & mixture & & & & & reaction rate & & - TAMS & ZAMS (R$_\odot$)\\
\hline
&&&&&&&&&\\
STAREVOL & AGSS09 & 1.973 $\odot$ & Allard+12 & Modified & OPAL & NACRE II & No & 53 Myr & 0.892 \\
This work & $Z=0.0134$  & & $\tau_{atm}=2$ & PTEH95 & F05 & & & - 8.77 Gyr &\\ 
& $Y=0.269$ &&&&&&&&\\
\hline
&&&&&&&&&\\
BHAC15 & Caffau+11 & 1.6 & Allard+12 & SCVH95 & OPAL & CF88 & No & 55 Myr& 0.898 \\
\cite{BHAC15} & $Z=0.0153$ & & Rajpurohit+13 & & AF94 & & & - 8.34 Gyr &\\
& $Y=0.28$ && $\tau_{atm}=100$ &&&&&&\\
\hline
&&&&&&&&\\
DSEP &  GS98 & 1.938 $\odot$ & Hauschildt+99 & CK95 & OPAL & Alderberger+98 & Yes & 55 Myr& 0.872 \\ 
\cite{Dotteretal2008} & $Z=0.0189$ & & $\tau_{atm}=\tau_{\rm eff}$ & FreeEOS4 & F05 & & & - 8.81 Gyr &\\
& $Y=0.274$ &&&&&&&&\\ 
\hline
&&&&&&&&&\\
STAREVOL & GN93 & 1.6 & P92+Eriksson94 & Modified & OPAL &CF88 & No & N/C &\\
\cite{Siess2000} & $Z=0.02$ & & +Kurucz91 & PTEH95 & AF94 & & & - N/C &\\
& $Y=0.28$ && $\tau_{atm}=10$ &&&&&&\\
\hline
&&&&&&&&&\\
FRANEC & AS05 & 1.68 $\odot$ & BH05 & OPAL06 & OPAL & NACRE & No & 56 Myr& 0.882\\
\cite{Tognelli2011} & $Z=0.01377$ & & $\tau_{atm}=10$ & & F05 & LUNA && - 9.13 Gyr &\\
& $Y=0.253$ &&&&&&&&\\
\hline
&&&&&&&&&\\
YREC & GS98 & 1.875 & Allard+11 & OPAL05 & OPAL & BP92 & No & 47 Myr& 0.735 \\
\cite{Spadaetal2011}& $Z=0.0163$ & & $\tau_{atm}=2/3$ & SCVH95 & F05 & & & - 8.16 Gyr &\\
& $Y=0.274$ &&&&&&&&\\
\hline
&&&&&&&&&\\
PARSEC & Caffau+09 & 1.74 & Allard+11 & FreeEOSv2.2.1 & OPAL & JINA REACLIB & Yes & 46 Myr& 0.876\\
\cite{Bressanetal2012}& $Z=0.014$ & 1.77 & $\tau_{atm}=2/3$ & & AESOPUS & & & - 8.06 Gyr &\\
& $Y=0.273$ &&&&&&&&\\
\hline
&&&&&&&&&\\
MESA & AGSS09 & 1.82 & ATLAS12 & OPAL + & OPAL & JINA REACLIB & Yes & 54.5 Myr& 0.888 \\
\cite{Choietal2016} & $Z=0.0142$ & & $\tau_{atm}=100$ & SCVH95 + & F05 & & & - 8.28 Gyr &\\
& $Y=0.2703$ &&& MDM12 &&&&&\\
\hline
&&&&&&&&&\\
CLES & GN93\footnote{Except light elements (Li, Be, B)} & 1.6 & Kurucz (1998) & OPAL & OPAL & NACRE & No & 49 Myr& 0.894\\
\cite{Montalban2008} & $Z=0.02$ & & $\tau_{atm}=2/3$ & & AF94 & & & - 8.57 Gyr &\\
& $Y=0.28$ &&&&&&&&\\
\multicolumn{10}{c}{}\\
\end{tabular}
\label{tab:parameter}
\caption{ZAMS : X$_c$=0.998X$_{c,i}$ and TAMS : X$_c$<0.002X$_{c,i}$. {\bf References :} \footnotesize AGSS09 : \cite{AsplundGrevesse2009},  Allard+12 : \cite{Allard2012}, OPAL : \cite{IglesiasRogers1996}, NACRE II : \cite{Xu2013a}, PTEH95 : \cite{Pols1995}, F05 : \cite{F05}, Caffau+11 : \cite{Caffau2011}, Allard+11 : \cite{Allard2011}, SCVH95 : \cite{SCVH95}, CF88 : \cite{CaughlanFowler88}, Rajpurohit+13 : \cite{Rajpurohit2013}, AF94 : \cite{AF94}, GS98 : \cite{GS98}, Hauschildt+99 : \cite{Hauschildt99}, CK95 : \cite{CK95}, Adelberger+98 : \cite{Adelbergeretal98}, FreeEOS(2,4) : \cite{Irwin12}, P92 : \cite{Plez1992}, Eriksson94 : Eriksson \citetext{priv. comm.} Kurucz91  : \cite{Kurucz1991}, AGS05 : \cite{AGS05}, BH05 : \cite{BH05}, OPAL06 : \cite{RN2002}, LUNA : \cite{LUNA2006a}, BP92 : \cite{BP92}, Caffau+09 : \cite{Caffau09}, JINA REACLIB : \cite{Cyburt10}, AESOPUS : \cite{MarigoAringer09}, ATLAS12 : \cite{Kurucz93}, MDM12 : \cite{MDM12}.}
\end{center}
\end{table}
\end{landscape}

\subsubsection*{PARSEC : \cite{Bressanetal2012}, \cite{Chenetal2014}}
When comparing our models with those computed by \cite{Bressanetal2012} and  \cite{Chenetal2014} with the PAdova and TRieste Stellar Evolution Code (PARSEC), we see a significantly different behavior in the HR diagram that cannot be attributed to their EOS which is very similar to ours. The large discrepancies at low temperature (lower mass models) is most likely due to the fact that PARSEC models use a very specific set of low-temperature opacities from the \emph{AESOPUS} tool. The Rosseland mean opacities provided by this tool are shown to differ the most from OPAL and \citet{F05} in this domain \citep{MarigoAringer09}. This reveals how difficult it is to determine what can actually be considered a suitable set of physical inputs for this phase of stellar evolution. 

\subsubsection*{MESA : \cite{Choietal2016}}
For comparison we use the Modules for Experiments in Stellar Astrophysics (MESA) Isochrones and Stellar Tracks (MIST) published by \cite{Choietal2016}. This grid is computed with standard physics adapted for solar-type stars and are  very close to ours. However, as with DSEP, MESA models account for overshoot at the interface of convective regions. In their case, they use an exponential diffusive overshoot \citep{Herwig2000} where the overshooting parameter $f$ is fixed at 0.016 for the core and 0.0174 for the envelope. Consequently they develop a more extended main sequence, as DSEP.

\subsubsection*{CLES : \cite{Montalban2008}}
A grid of models was computed with the Code Li\`egeois d'\'Evolution Stellaire (CL\'ES) for the analysis of  CoRoT data and compared to other evolutionary codes not presented here (see \citealt{Montalban2008} and references therein). The differences observed in Fig.~\ref{Fig:allhrd} along the PMS are due to the different boundary (atmosphere) conditions. Then both sets of tracks converge on the MS, except for the most massive models. In particular, the hook observed at the end of the main sequence of the 1.2\Ms is \emph{not} due to any overshooting but to the higher Z associated to \citeauthor{GN93}'s abundances used in CLES models. This higher metallicity, by increasing the opacity, favors the development of a CC at lower masses as in the early STAREVOL grid from \cite{Siess2000} where the Sun had, for a short period of time, a very small CC on the MS.

\subsubsection*{Global comparison of the PMS lifetime and ZAMS radius}
The last two columns of Table~\ref{tab:parameter} give the PMS and MS lifetime and the ZAMS radius\footnote{We arbitrarily define the ZAMS as the time when 0.2 percent of the initial hydrogen has been burnt at the center.}  of the solar-like models. 
The PMS duration clusters around two values, one at 55 Myr with a dispersion of 2 Myr, and another one at 47 Myr with the same dispersion. We investigated several trails to interpret such behaviour looking for the effect of differences in the initial chemical composition, starting point on the Hayashi line and initial central temperature, numerical timestep or deuterium burning rate. None of these quantities lead to a conclusive trend but the numerics of the code, in particular the discretisation and shell rezoning can have a noticeable effect that was reported e.g. in core helium burning or AGB stars \citep{Siess2002}. The terminal age main sequence varies between 8.06 Gyr (PARSEC) and 9.13 Gyr (FRANEC), with no specific trend nor clustering of ages depending on the input physics. We may just emphasize the puzzling result concerning the YREC and PARSEC models, which differ in almost every physical paramters but present very similar ages at both the ZAMS and TAMS.
Except for the YREC models, the radii seem to be all consistent with a ZAMS radius of $0.888 \pm 0.015$, regardless the age of the ZAMS.

This comparison sheds light on the heterogeneity of the stellar evolution models predictions for a given initial mass and ``solar metallicity''. This should be kept in mind whenever various stellar evolution models  are combined or used to interpret observational data.

\section{Angular momentum evolution}\label{sec:AM}

After this comparison with the other standard PMS models available in the literature, we now turn to the specificity of our work, namely the effect of rotation. In this section, we explore in details the rotational behavior of our models. 
First, we compare our predictions to some characteristics of our standard models. Second, we discuss the behaviour of the surface rotation of our models as a function of mass and age.
Third, we compare our predictions to observed surface rotation periods at $Z_\odot$. 
Finally, we present a thorough analysis of the internal transport of AM as a function of mass, metallicity, and age.

\subsection{Effect of rotation on the evolution in the HRD }
\label{sec:DL}

\begin{figure}
\includegraphics[width=0.49\textwidth]{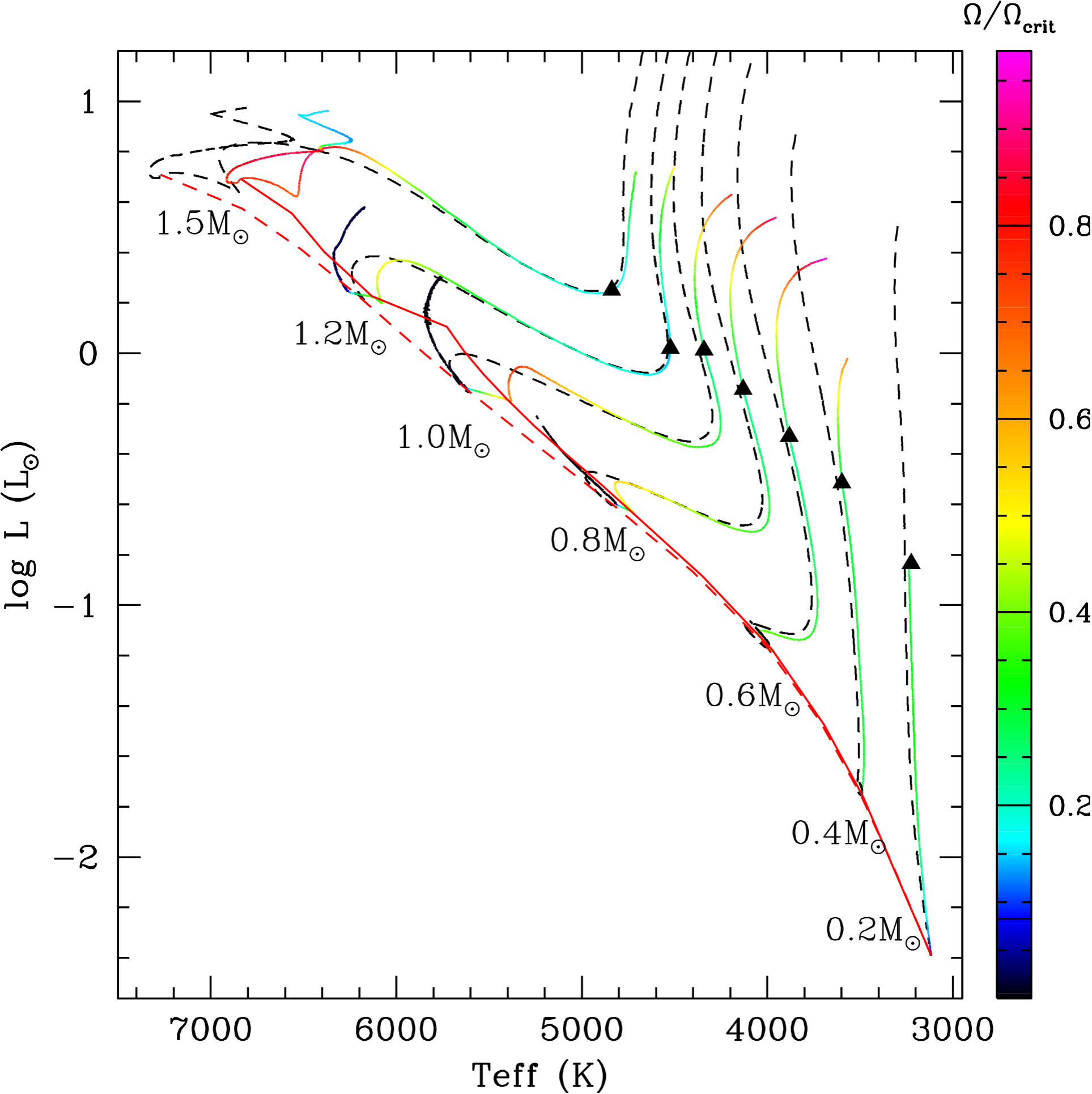}
\caption{HR diagram of solar metallicity models without (dashed black line) and with rotation (solid colored lines; here we show the fast rotators). 
The values of the surface velocity normalized to the break-up value ($\Omega/\Omega_\mathrm{crit}$) increase from blue to red as shown on the right color bar. The black triangles indicate when the rotating models are released from their disc. The red lines indicate the standard (dashed) and rotating (solid) ZAMS.}
\label{Fig:hrd_rothydro}
\end{figure}

\begin{figure}
\includegraphics[width=0.49\textwidth]{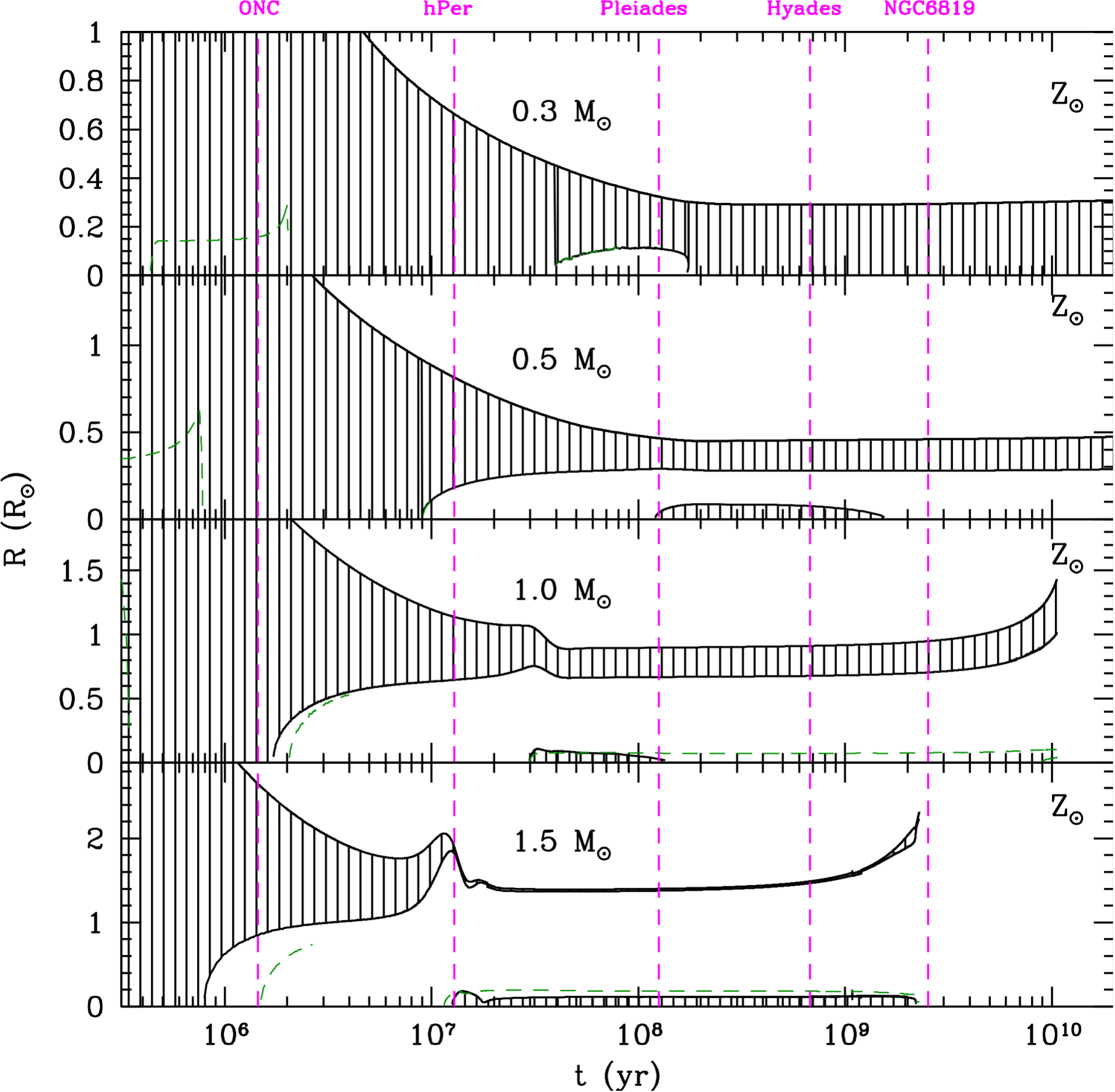}
\caption{Kippenhahn diagram showing the evolution of the internal structure of the non-rotating solar metallicity models of 0.3 (top), 0.5, 1.0 and 1.5\Ms (bottom) from the PMS up to the end of the main sequence. The upper line represents the surface radius and hatched areas refer to convective regions. The green line displays the H-burning limit. The five pink vertical lines indicate the ages of open clusters used as markers of the evolution.}
\label{Fig:kipdiagM}
\end{figure}

\begin{figure*}
\includegraphics[width=\textwidth]{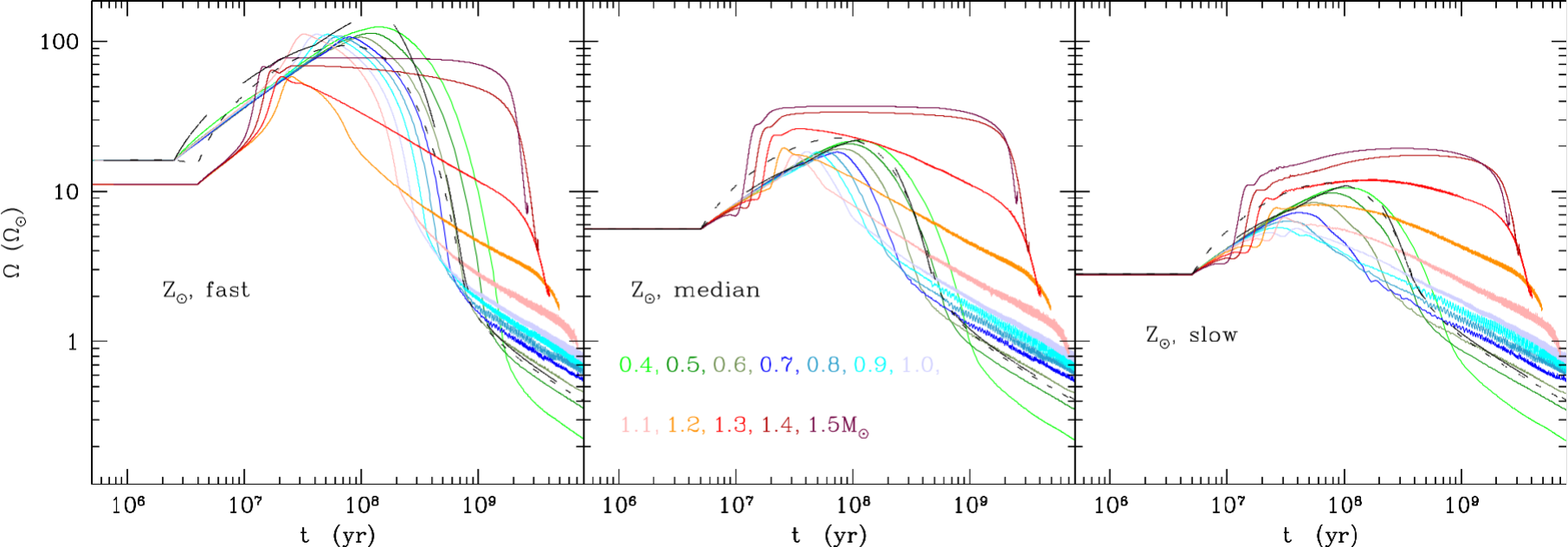}
\caption{Evolution of surface angular velocity as a function of time at $Z=Z_\odot$ for fast (left), median (center) and slow (right) rotators from 0.4$M_\odot$ (neon green) up to 1.5$M_\odot$ (burgundy), the 0.2 and 0.3$M_\odot$ are shown in black short- and long-dashed lines, respectively.}
\label{Fig:vsurf}
\end{figure*}

Figure~\ref{Fig:hrd_rothydro} compares the evolutionary tracks in the HRD of selected standard and fast rotating models at solar metallicity. The colours indicate the surface angular velocity normalized to the break-up angular velocity. 
As shown by \cite{ES76}, the deformation of the stellar structure by the action of centrifugal forces is expected to shift the track of a rotating star in the HR diagram toward lower effective temperatures. Indeed, in case of fast rotation, the radius is larger, the equator cooler and the mean effective temperature of the star is thus lower.

For the mass range considered in the present grid, this effect is only relevant for the fast rotating models. The median and slow rotators follow the same evolutionary path in the HR diagram as their standard counterparts.

This shift towards lower temperatures in the evolutionary tracks of fast rotators is visible at different locations on the PMS and MS depending on the initial mass : 
1) at the tip of the Hayashi line in the HR diagram (red part of the tracks in Fig.~\ref{Fig:hrd_rothydro}) where the model stars are initially very extended and contracting very rapidly; 2) at the end of the PMS for models more massive than 0.6 $M_\odot$, that undergo a final contraction after ignition of core nuclear reactions, before they arrive on the MS; 3) during the MS evolution for the 1.4 and 1.5 $M_\odot$ models.

As can be seen in Fig.~\ref{Fig:hrd_rothydro}, the ZAMS of the fast rotators is reached at cooler temperatures due to the effects of the centrifugal acceleration, and this shift increases with initial mass\footnote{The ZAMS of the massive rotating models moves closer to the standard location due to the smaller initial angular velocity assumed for the fast rotating 1.3 to 1.5 M$_\odot$ models.}. The lower the initial mass the closer to the standard location of the ZAMS. 

Below 0.6 $M_\odot$, the ratio $\Omega/\Omega_\mathrm{crit}$ never exceeds 0.4 after the star is decoupled from its disc (indicated with black triangles in Fig.~\ref{Fig:hrd_rothydro}). The deformation of the stellar structure by centrifugal forces is negligible and the rotating tracks on the HR diagram follow the standard ones.\\

Between 0.6 $M_\odot$ and 1.3 M$_\odot$ at solar metallicity, the models reach fairly high rotation rates on their arrival on the ZAMS (up to $0.9 \Omega_{\rm crit}$) and in the HR diagram they thus appear much cooler. However, owing to their thick convective envelope, they are efficiently spun down, and converge towards the standard non-rotating tracks on the MS.
On the early MS, while the star is almost still in the HR diagram, its surface velocity can change substantially 
\citep[\eg][]{Barnes2016}. For example, it takes $2\times 10^8$ years for a fast rotating 1\Ms model to spin down from 75\% to less than 10 \% of the critical velocity, while in the same amount of time the luminosity increases by only 1-2\% and less than 2\%  of hydrogen has been burnt in the core.
This rapid spin down leads to an increase of the effective temperature at almost constant luminosity from the fast rotating cooler ZAMS to the slow rotating hotter MS. \\ 
The 1.4 and 1.5 M$_\odot$ models have a very thin convective envelope on the MS, and hence lose almost no AM through magnetic braking. They maintain a high $\Omega/\Omega_\mathrm{crit}$ value during most of the MS, so their evolutionary tracks in the HRD remain cooler than the standard ones.

\subsection{Evolution of surface rotation on PMS and MS}

The evolution of the surface rotation of low-mass stars during the PMS and MS is due to the combined effects of the structural changes, of the efficiency of the torque exerted by magnetized winds at the stellar surface, and of the internal transport of AM.
Figure~\ref{Fig:vsurf} presents the evolution of the surface angular velocity of the fast (left), median (center) and slow (right) rotating models of all masses at solar metallicity. 

On the PMS, as long as the star is coupled to its disc, i.e. its angular velocity kept constant in the model, the break-up velocity increases when the star contracts. The ratio $\Omega/\Omega_\mathrm{crit}$ thus decreases over this period so all the rotating models progressively join their standard tracks (see Fig.~\ref{Fig:hrd_rothydro}).

After the star-disc decoupling, the stars are free to spin up, and reach a maximum velocity that is larger for higher stellar mass. This surface acceleration is driven by the structural changes. In the case of the initially fast rotators, the most massive models can even reach close to break-up surface velocities as they approach the ZAMS (red part of the tracks on Fig.~\ref{Fig:hrd_rothydro}).

All the models with masses below 1.4\Ms (at $Z_\odot$) reach their maximum velocity at their arrival on the ZAMS and then spin down on the MS when magnetic braking kicks in (see Fig.~\ref{Fig:vsurf}). This peak velocity coincides with the onset of core convection following the activation of the $^{12}$C(p$,\gamma$) reaction that stops the star's contraction. The fully convective 0.2 and 0.3 $M_\odot$ models start spinning down when the contraction rate has slowed down and the magnetized wind torque has strengthened (around $10^8$~yr for the 0.3\Ms). 

In the fastest rotators, the magnetic field is saturated ($Ro$ < 0.14)  when the effect of the stellar wind torque first becomes effective, and then switches to the unsaturated regime as the surface angular velocity decreases. The early MS evolution of the surface velocity of all fast rotators thus starts with a rapid spin down followed by a more progressive decline decrease in the spin rate. This transition between the saturated and unsaturated regime is marked in the Fig.~\ref{Fig:vsurf} by the change in the slope. We also notice that in the unsaturated regime the spin velocity follows a Skumanich-like relation with $\Omega \propto t^{-p}$. Finally, the slow and median rotators with masses ($M \ge 0.9M_\odot$)  and the fast rotators with $M \ge 1.3M_\odot$ always evolve in the unsaturated regime. 

The magnetic braking as included in our models however proves to be inefficient for the most massive models ( $\geq$ 1.4 M$_\odot$). These stars have a very thin convective envelope with a high convective turnover timescale (\ie a high Rossby number), on the MS (see Fig.~\ref{Fig:kipdiagM}), and hence lose almost no AM through magnetic torques. The observations also become very sparse in this mass range due to the lack of surface magnetic spots in such stars, which are needed to consistently retrieve the rotation period from photometry. 

In their late evolution, the surface velocities of models with the same initial mass but different initial rotation rate converge to the same value, so no constraints can be obtained on the initial AM content of stars based on their MS rotation rate \citep[\emph{see also}][]{Kawaler88,Amard2016}.\\

The overall behavior described hereabove is compatible with the observational results by \citet{Folsom2016,Folsom2018} who showed that the evolution of the magnetic field strength and of its geometry - which define the torque applied at the stellar surface - are primarily driven by structural changes during the PMS while on the MS, they correlate with the angular velocity of the star. 

\begin{figure*}
\includegraphics[width=\textwidth]{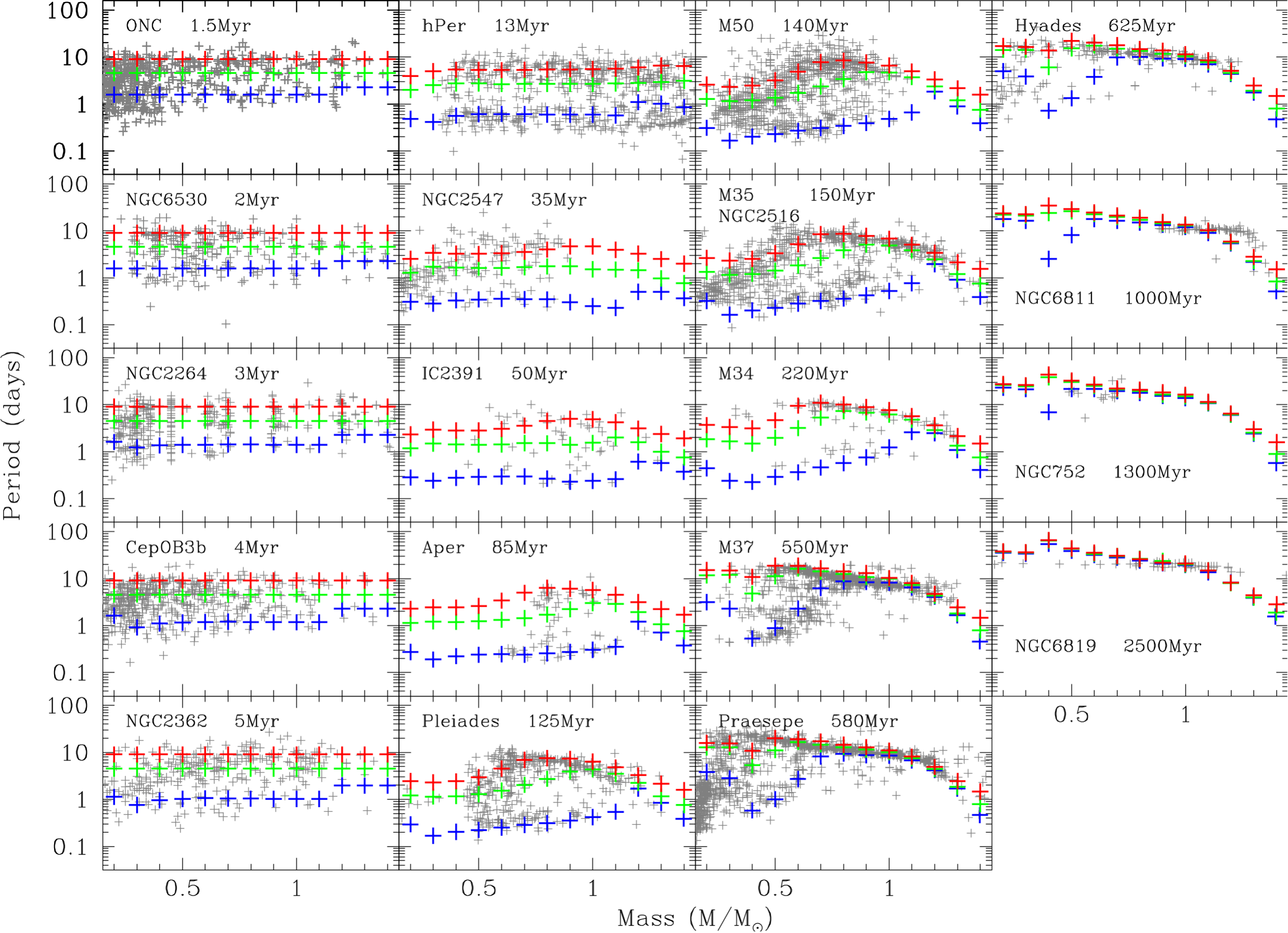}
\caption{Comparison over the whole mass range, between 0.2 to 1.5\Ms, of the rotation period distributions of our solar metallicity models with observations from open clusters of increasing ages  \citep[grey crosses, data from ][]{GB13,BouvierPPVI,Douglasetal2016,Douglasetal2017,Aguerosetal2018}. 
The red, green and blue crosses represent the rotation periods of the slow, median and fast rotating grids, respectively. Masses and ages of clusters members are taken from the literature.}
\label{Fig:compobs0210}
\end{figure*}

\subsection{Surface rotation - comparison to observations}

In \citet{Amard2016}, we compared the surface angular velocity evolution predicted by the 1\Ms models at $Z_\odot$ to rotation periods measurements of solar-type stars in star forming regions and young open clusters (1Myr - 2.5Gyr). 
The models were computed with different physical descriptions for the internal transport and extraction of AM. We found an overall good agreement between the predicted and observed surface rotation rate, with the models presenting a relatively strong differential rotation profile along most of the evolution. 
We concluded that the rotational evolution of young stars is insufficient to constrain the internal transport of AM.

In this section, we extend this comparison to a broader range of stellar masses for models with updated input physics. We focus on solar metallicity where more data are available and allow to cover a larger range of ages. We recall that each grid, characterized by its metallicity and initial angular velocity $\Omega_{init}$, is computed with the same value for the disc-coupling timescale ($\tau_{DL}$) independently of the initial stellar mass and metallicity (except for the fast rotating models with $M \in [1.2-1.5]$\Ms, see Table~\ref{tab:rot} and \S~\ref{sec:extAM}). 
Figure~\ref{Fig:compobs0210} shows a comparison of the surface rotation periods predicted by our solar metallicity grid to the observed rotation periods of open clusters members, at the age given in the literature for each cluster.
The overall shape and evolution of the observed rotation period as a function of mass is well reproduced by our models. 
Here we summarize the main observational points and compare them to model predictions.

\begin{itemize}
\item During the first few million years, the rotation period presents a large dispersion ($\Delta P_\mathrm{rot} \approx 10$ days) that remains roughly constant (see e.g., ONC, NGC6530, NGC2254, CepOB3b and NGC2362, first column of Fig.~\ref{Fig:compobs0210}). 
Nonetheless, \cite{Somersetal2017} mention the presence in young clusters of a correlation between the stellar mass and the rotation period, with the less massive stars having the shortest period. This may indicate that the less massive stars are already spinning up and therefore could have shorter disc lifetimes. We did not account for this feature but despite this limitation our models still remain in fair agreement with observations at these very early ages.

\item The second phase (second column of Fig.~\ref{Fig:compobs0210}) corresponds to the time when the PMS stars are released from their disc and free to spin up. For clusters covering this period (a few $10^6$~yr), the dichotomy between fast and slow rotators sequences is very clear, as exemplified by hPer (13 Myr).
Some observed stars are really close to the break-up velocity and still, they are not expected to have ended their contraction. With our adopted initial conditions, we are able to reproduce most of the  spread in rotation period in hPer and the two sequences running along the red and blue crosses observed in pre-ZAMS clusters.

\item By the age of the Pleiades (125 Myr), the models above 1.2\Ms have been efficiently braked and the initially slow and fast rotators start to merge into a unique sequence. This is not the case for the lower mass models that evolve more slowly and may still be contracting. 

\item In the third column of Fig.~\ref{Fig:compobs0210}, we see a variation of the observed dispersions of slow rotators with mass and age. Stars with a lower mass reach this sequence later than their more massive counterparts because their contraction phase lasts longer,
and because their magnetic field saturates for a lower rotation rate, they enter a regime of saturated magnetic field for a longer time which delays their spin-down.
The models are also able to reproduce the progressive convergence of the slow (red) and fast (blue) sequences.
At the age of Praesepe (580 Myr), the fit to the observed dispersion is very good down to 0.4\Ms, but the models fail to reproduce the short rotation period of the less massive stars. This discrepancy between models predictions and observations has been discussed in \cite{Aguerosetal2018} and  appears at the mass  transition where the star remains fully convective.
For these very low-mass stars, our braking prescription is too efficient and/or happens too early. This indicates that the expression and calibration of the braking law should be modified in this low-mass fully convective regime \citep[e.g.][]{Mattetal2015}.

\item We then reach the fourth column where the data can be used for gyrochronology. 
By 1 Gyr, all stars have spun down. Our models can reproduce fairly well the evolution of the rotational velocity of solar-type stars but they fail to account for the relatively flat distribution of periods over the entire mass range. Above 1.2$M_\odot$, the predicted rotational period is too short compared to the observations and at smaller masses, the discrepancy is not as severe but our models slightly overestimate the spin rate. We point out that above 1.2$M_\odot$, main sequence stars have a thinner convective envelope than their lower mass counterparts and also develop a convective core during central H-burning. The structure of the dynamo-generated magnetic field may change with the size of the convective envelope, going from a dominant dipolar large scale component to a more multipolar field organized on smaller scales \citep{Donati2011}.
This would surely affect the braking efficiency, even if it is not clear whether such an evolution would explain the observed discrepancy. Indeed, a field organized as a higher degree multipole is expected to have a weaker lever arm, and hence reduce AM loss \citep[e.g.][]{Reville2015a,FinleyMatt2018,See2018,Garraffo2018}, which is opposite to what appears to be needed to reconcile our models with observations. We finally note that the 0.4 and 0.5\Ms models are spinning too slowly compared to the observed rotation periods in NGC 6819. It likely comes from the incomplete transport of AM and the corresponding calibration constant ($K$) that we selected for the 1.0\Ms models. These stars are on the verge of the fully convective mass domain and have a very deep convective envelope, thus they are rotating nearly as solid bodies (since we assumed constant angular velocity in convective regions). For example, if we had considered a solid body for the Sun, the calibration constant would have been smaller, resulting in a smaller torque and a larger angular velocity at later ages (see alo \cite{Amard2016} for discussion on the impact of the constant $K$).

\end{itemize}

\subsubsection*{Cluster age uncertainties}
The cluster ages reported in Fig.~\ref{Fig:compobs0210} are taken from the literature\footnote{We actually plan to redetermine the cluster ages with our own isochrones in a future paper.}, and the masses for the sample stars are from \citet{GB13,BouvierPPVI,Douglasetal2016,Douglasetal2017,Aguerosetal2018} and references therein. \\
The ages of the youngest clusters (up to hPer) are relatively uncertain with sometimes a factor of two uncertainty depending on the sets of isochrones used to fit their color-magnitude diagram (CMD). One of the main reasons for this uncertainty is the poor radius determination of very low-mass and very cool dwarf models.
Indeed, eclipsing low-mass binaries exhibit inflated radii in comparison to the ones provided by any evolutionary models, which impacts their location in the HR diagram \citep[see \eg][]{BHAC15}. 
\cite{Belletal2013} provided empirical corrections to theoretical isochrones in order to better reproduce the colour-magnitude diagram in all colours. These corrections give ages up to a factor of 2 greater than the ones obtained with standard isochrones. 
However a big caveat of these corrections is that, except for the age, all the other parameters of the corresponding evolutionary models are not consistent anymore. 
\cite{SP2015} proposed that stars populating the youngest open clusters are strongly magnetized and would develop a high activity leading to a high spot coverage. These cool spots on the surface would then induce a back-reaction on the structure and the star would puff up and mimic the expected inflated radius.
Finally, \cite{FeidenChaboyer2012} provided some evolutionary models including a simplified treatment of the effects of magnetic field  on the structure. This formalism leads to a less efficient convection that inflates the stellar radius and reproduces fairly well the CMD of young open clusters but requires very strong magnetic fields.

\subsection{Internal rotation - Effect of initial mass}

\begin{center}
\begin{figure}
\includegraphics[width=0.48\textwidth]{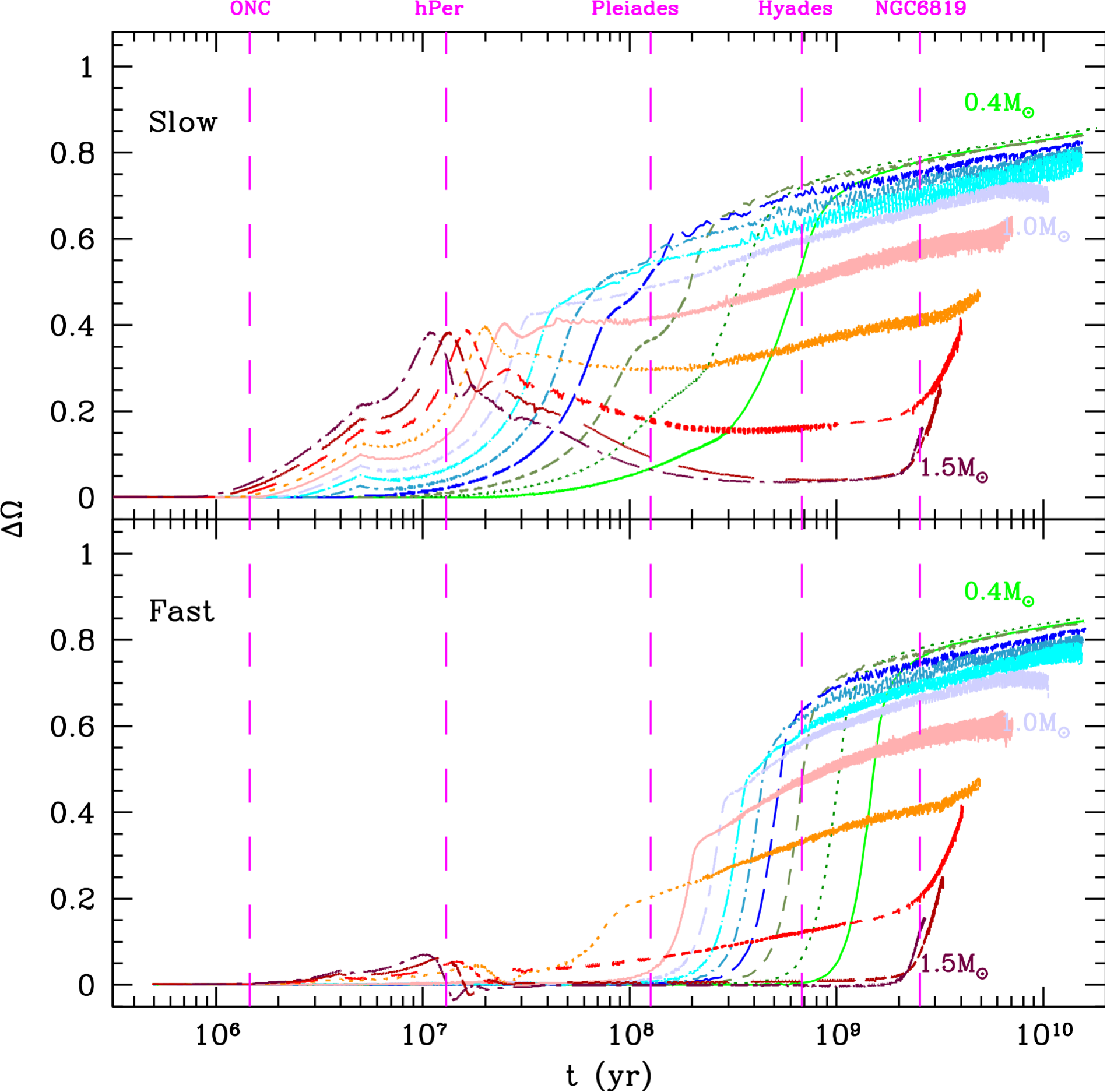}
\caption{Differential rotation as a function of time for slow (top) and fast (bottom) rotators at $Z = Z_\odot$. The color code is the same as in Fig.~\ref{Fig:vsurf}.}
\label{Fig:HRdiffrot}
\end{figure}
\end{center}

\begin{figure}
\includegraphics[width=0.48\textwidth]{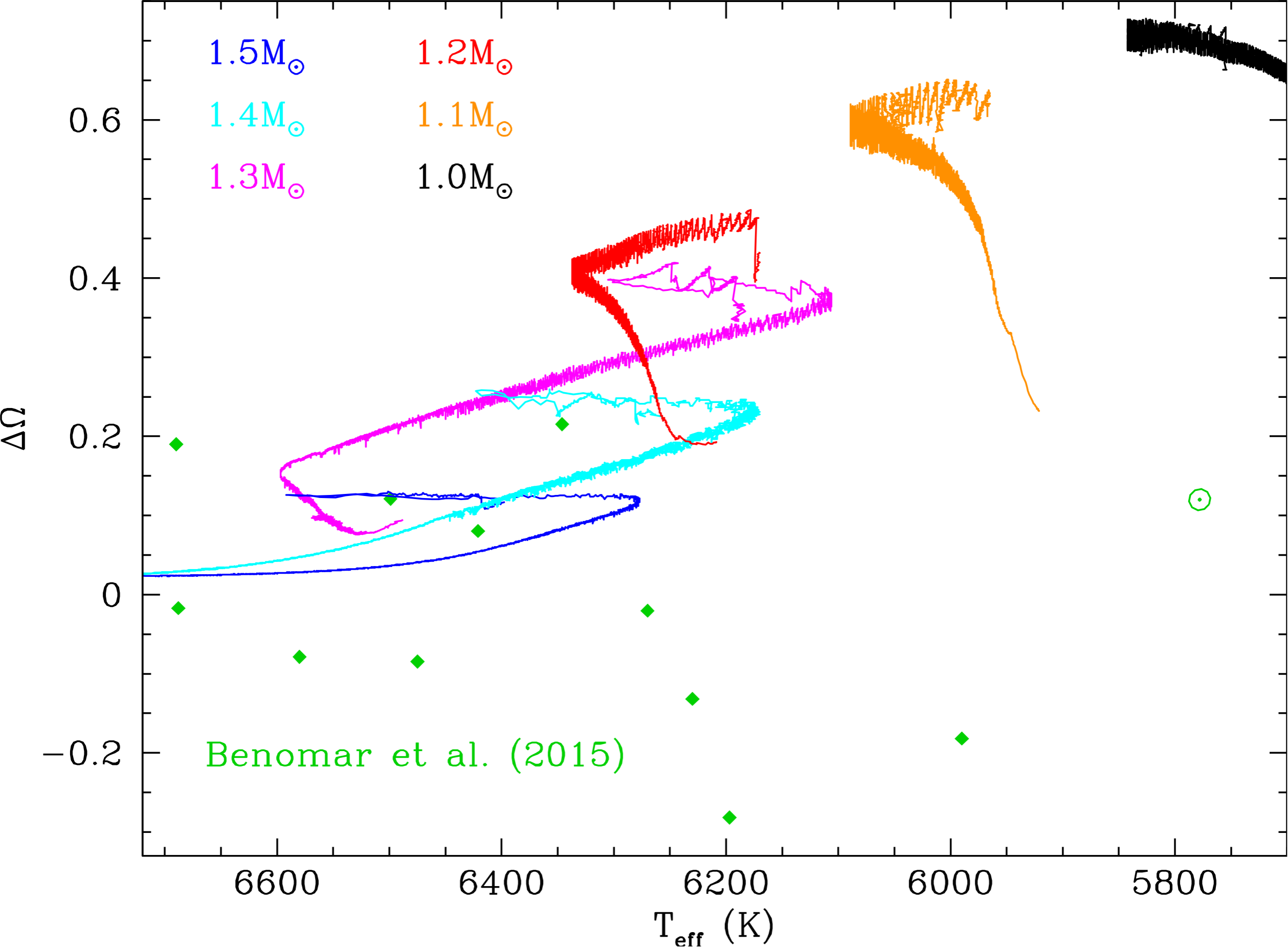}
\caption{Differential rotation ($\Delta\Omega$) as a function of effective temperature for our solar metallicity 1.0 to 1.5 M$_\odot$ models with a median initial rotation rate. Green diamonds show the value of $\Delta\Omega$ from \citet{Benomar2015}'s data. The solar value as given by \citet{Fossat2017} is indicated by $\odot$.}
\label{Fig:diffrotobs}
\end{figure}

Figure~\ref{Fig:HRdiffrot} shows the level of internal differential rotation $\Delta \Omega$ for the slow and fast rotators of all solar metallicity models\footnote{0.2 M$_\odot$ and 0.3 M$_\odot$ models are not presented as they evolve as fully convective stars.} as a function of time from the onset of the radiative core to the TAMS (or up to 15 Gyr for the models that have a longer MS lifetime). We express it as :
\begin{align}
\Delta\Omega = \frac{\Omega_C - \Omega_S}{\Omega_C + \Omega_S} && \textrm{with} && \Omega_C=\int_0^{M_{CZ}}\Omega dm
\label{deltaomega}
\end{align}
where $M_{CZ}$ is the mass coordinate at the base of the convective envelope and $\Omega_S$ the surface angular velocity. With this formulation, $\Delta\Omega \rightarrow -1$ corresponds to a slow-rotating core with a fast rotating envelope, $\Delta\Omega = 0$ a flattened rotation profile on average, and $\Delta\Omega = 1$ is characteristic of a fast-rotating core with a slowly rotating surface.
Note that $\Omega_C$ is not comparable to the solar core value derived by helioseismology, as in \eg \citet{Fossat2017} where they claim that the solar core is rotating five times faster than the solar surface. 
According to our unit system, $\Delta\Omega_\odot = 0.12$ (see later in this section Fig.~\ref{Fig:diffrotobs}).
As seen in Fig.~\ref{Fig:HRdiffrot}, all our models evolve between these last two cases, namely $\Delta\Omega = 0$ and $\Delta\Omega = 1$.

During the PMS phase, the contraction of the star, and then the appearance of the convective core (when it exists), generate a strong meridional circulation which remains the main driver for AM transport. Meridional circulation in these models only transports AM from the core to the surface. This is in agreement with our previous study of solar-mass solar metallicity stars in \cite{Amard2016}. The efficiency of the circulation depends directly on the rotation rate. Therefore, the more rapidly the star is spinning, the closer to solid-body it is. This is valid for all the stellar masses we consider here.

In slow rotating models, $\Delta\Omega$ increases as the radiative core appears during the PMS as shown on the top panel of Fig.~\ref{Fig:HRdiffrot}. 
Its value rises from 0 at the age of the ONC (the fully convective star is in solid-body rotation), up to $\Delta\Omega = 0.7$ at the age of the Hyades for the 0.5\Ms  and $\Delta\Omega = 0.4$ at the age of hPer for the slow rotating 1.4\Ms model. This strong differential rotation results almost exclusively from the structural changes (stellar contraction and shrinkage of the convective envelope) because at that stage, the rotation rate is slow and the internal AM transport by meridional circulation and shear turbulence is negligible.\\
Then on the MS, we can distinguish two families of slow rotators.  Models with M$_{ini} > 1.2 M_\odot$ have a thin convective envelope (see Fig.~\ref{Fig:kipdiagM}) characterized by a short convective turnover timescale so, for a given rotation rate, they are associated with a high Rossby number (see Sect.~\ref{sect:Rossby}). They are thus expected to have a less active dynamo, and the torque applied at their surface is reduced. This implies that more massive models can maintain a high rotation rate during their main sequence evolution which in turn can trigger stronger meridional currents capable of reducing the degree of differential rotation.\\ 
For stars with M$_{ini} < 1.2 M_\odot$ the differential rotation increases with time because they have more extended CE and can generate stronger magnetic torques. Their surface spin rate is thus lower and angular momentum transport redistribution in the radiative interior less efficient. A situation is thus reached in which the differential rotation rate keeps slowly  increasing due to the surface braking and the negligible effect of meridional currents.

The fast rotators present a very different behavior. They strongly couple their radiative core to their convective envelope for a longer period of time, that extends beyond $10^8$ yr. 
They have very strong meridional currents that carry AM from the radiative core to the convective envelope and reduce the differential rotation as discussed for the 1\Ms case in \cite{Amard2016}. 
When the stars are sufficiently spun down by the magnetized stellar wind, the surface angular velocity decreases, and differential rotation develops below the convective envelope where a nearly flat rotational profile was established during the fast rotating phase. If the convective envelope is too small to ensure an efficient braking, a flat rotation profile is maintained as can be seen for the 1.4-1.5\Ms. For these last 2 models, the sudden rise in $\Delta \Omega$ at the very end of the MS is due to the deepening of the surface convection zone.\\ 

\begin{figure}
\includegraphics[width=0.48\textwidth]{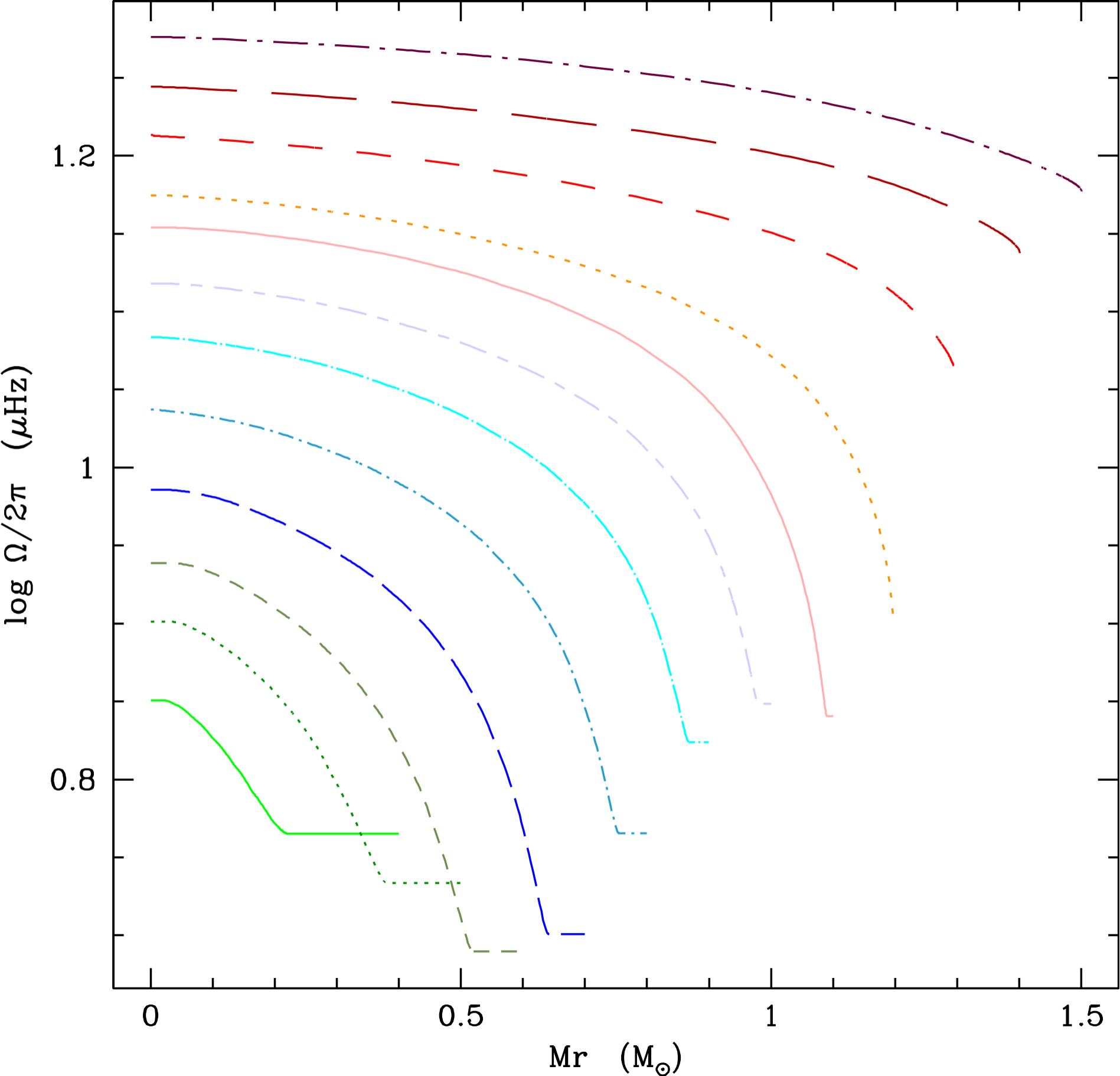}
\caption{Angular velocity profile as a function of the relative mass fraction of the median rotating models at solar metallicity for the 0.4-1.5\Ms mass range at the ZAMS. The color code is the same as in Fig.~\ref{Fig:vsurf}.}
\label{Fig:profomega_ZAMS}
\end{figure}

In Fig.~\ref{Fig:profomega_ZAMS}, the rotation profiles at the ZAMS of the median rotating models present a minimum surface angular velocity around 0.6\Ms (short-dashed olive green track). This is also observed with the slow and fast rotators around the same mass. Above this limit, the stars are braked less efficiently due to a smaller convective envelope, while below this limit, stars have been contracting efficiently towards the ZAMS, maintaining a higher surface rotation rate. 

To date, there are very few main sequence low-mass stars for which estimates of the core angular velocity is accessible through asteroseismic analysis. \citet{Benomar2015} published a sample of 22 F-stars with surface (envelope) and core rotation rates. We selected half of their sample, keeping those with [Fe/H] = $\pm$0.1 for which we computed $\Delta\Omega$ assuming a solid-body rotating radiative core, which is debatable. Fig.~\ref{Fig:diffrotobs} shows the obtained values as a function of effective temperature together with our solar metallicity models of equivalent masses. The solar value deduced from the -- controversial -- \citet{Fossat2017} rotation profile (see \citet{Schunker2018}) is also represented on this plot. 
The 1.4 and 1.5~M$_\odot$ models have a degree of differential rotation close to what is given by asteroseismology for T$_\mathrm{eff}$ > 6300K.
However, our models fail to reproduce lower temperatures data as our formalism does not produce any reversed rotation profiles -- with a core rotating slower than the surface. Internal gravity waves (IGWs) have been shown to produce this type of rotation profile and start to operate in this range of temperature \citep[\eg][]{Charbonnel2013}. A more in-depth study on that topic would therefore be a natural extension to this preliminary work. We also note that in our solar mass model, the coupling between the radiative interior and convective envelope is too weak to match the solar value derived from \citet{Fossat2017} data. A stronger coupling could however be achieved by the action of IGWs \citep[\eg][]{CharbonnelTalon2005Science}. Let us emphasize here that despite this discrepancy, our models are able to reproduce the only sound available observational constraint on the rotation of low mass stars which is given by the evolution of their surface rotation period. 

\begin{figure}
\includegraphics[width=0.48\textwidth]{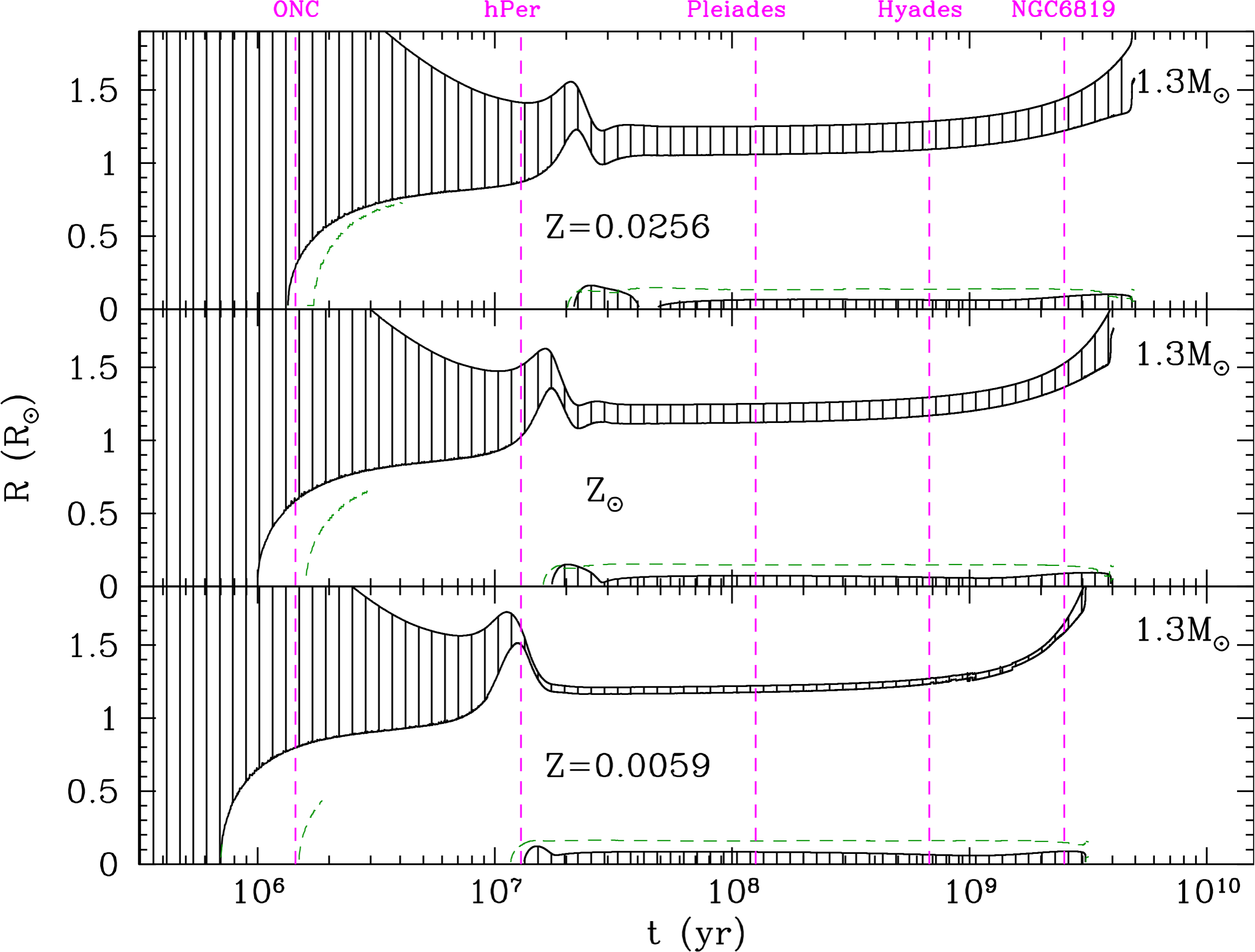}
\caption{Kippenhahn diagram of a standard 1.3\Ms models at three metallicities: $Z=0.02564$ (top), $Z_\odot$ (middle) and $Z=0.0059$ (bottom) from the PMS up to the end of the main sequence. The legend is the same as Fig~\ref{Fig:kipdiagM}.}
\label{Fig:kipdiagZ}
\end{figure}

\begin{figure*}
\includegraphics[width=\textwidth]{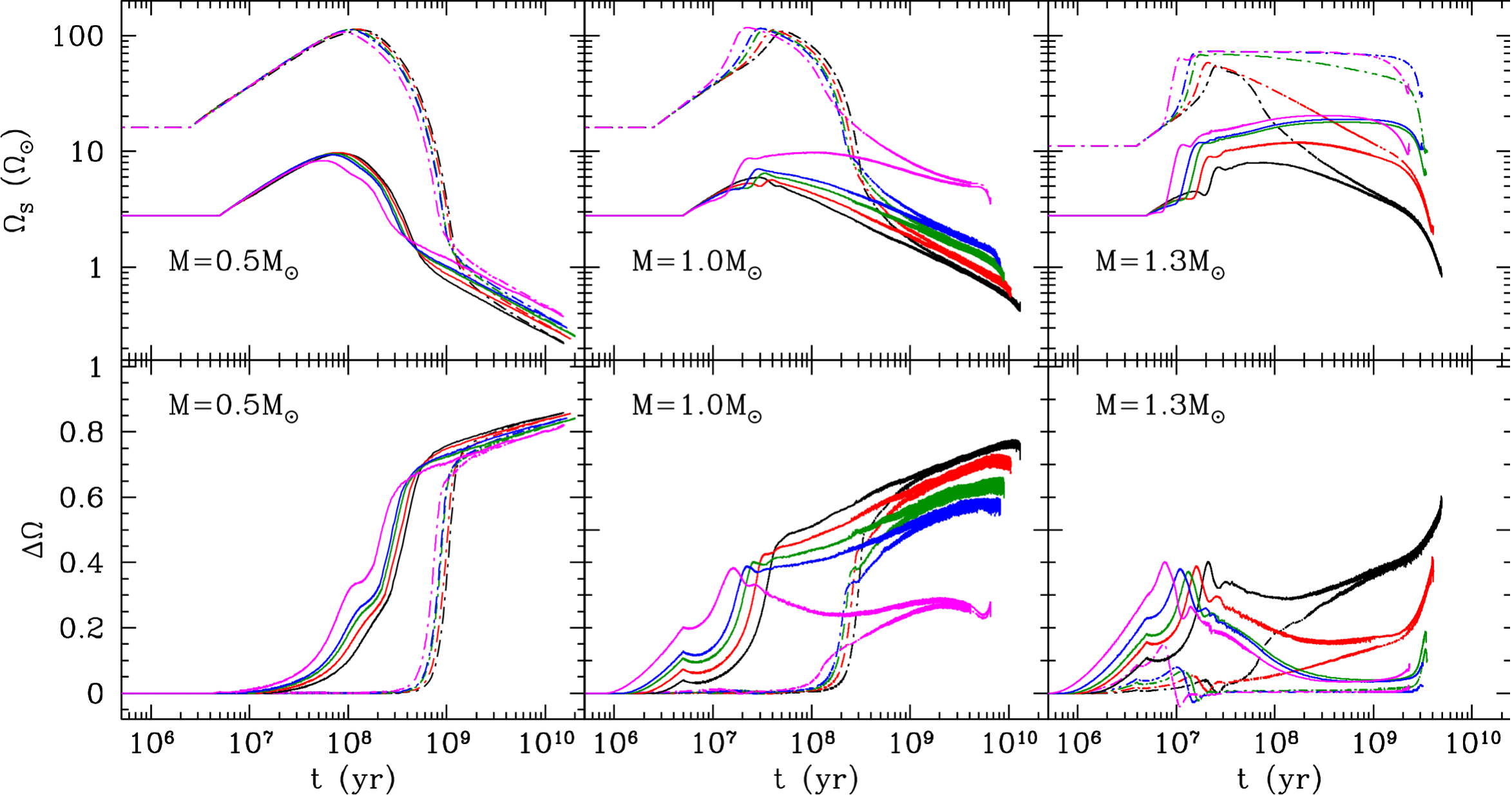}
\caption{{\bf Top.} Evolution of the surface rotation rate for three different masses at five metallicities, $Z=0.0022$ (magenta), $Z=0.0059$ (blue), $Z=0.0079$ (green), $Z=0.0134$ (red) and $Z=0.0256$ (black) for the fast (dashed-dot) and slow (solid) rotating cases.
{\bf Bottom.} Evolution of the relative differential rotation rate $\Delta\Omega$ using the same color coding as before. }
\label{Fig:rotation_allZ}
\end{figure*}

\subsection{Impact of metallicity on rotation}
The structure of a star depends on its mass but also on its chemical composition as illustrated in Fig.~\ref{Fig:kipdiagZ} showing the Kippenhahn diagram of a non-rotating 1.3\Ms model at three different metallicities.

For a given mass, a lower metal content reduces the global opacity, making the star hotter, more compact, and with a thinner convective envelope.
So when it comes to AM evolution, a lower metallicity generates a weaker torque so a larger surface velocity can be reached. Reciprocally, a higher metal content will produce slower rotators.
Additionally and as can be seen in Fig.~\ref{Fig:kipdiagZ}, metal poorer stars contract on a shorter timescale and their radiative core develops earlier on the PMS. Hence, they spin up more rapidly and reach the less efficient braking (saturated) regime earlier. 

In the top panel of Fig.~\ref{Fig:rotation_allZ}, we show the surface rotation rate for three masses, five metallicities and two initial velocities corresponding to the fast and slow rotators. The models with $Z=0.0059$ ([Fe/H]=-0.5 in blue) or $Z=0.0022$ ([Fe/H]=-1.0 in magenta) spin up faster than the ones with a solar or higher metal content ($Z=0.0134$ or $Z=0.026$ ([Fe/H]=+0.3)) and remains on the MS with faster surface rotation rates. 
The main difference is the transition to the unsaturated regime that is reached at higher velocities for lower metallicities models. For example, in the fast rotating 1\Ms case, the most metal poor models saturate around 30 $\Omega_\odot$ while the solar metallicity ones saturate at only 6 $\Omega_\odot$. Then on the unsaturated regime, the models converge to the same $\Omega \propto t^{-p}$ relation, independently of the metallicity.

Regarding the internal rotation properties, a metal-poor star as a more extended radiative region at a given evolutionary point on the MS, thus according to Eq.~\ref{eq:Fshear} and \ref{eq:Fcirc}, both the meridional circulation and shear turbulence AM flux are enhanced, leading to less differential rotation. 

This configuration favors solid-body rotation in metal-poor stars. This is illustrated in the bottom panel of Fig.~\ref{Fig:rotation_allZ} showing the evolution of $\Delta\Omega$ as given by Eq.~\ref{deltaomega}. 
For the three considered masses, the degree of differential rotation on the main sequence is always smaller for lower metallicity models. The result is especially clear in the case of the 1.3\Ms model, for which the evolution of the rotation velocity on the main sequence is strongly metal-dependent. 

Therefore, given an initial mass and rotational period, a lower metallicity model will reach a higher surface angular velocity and have less internal differential rotation. As a word of caution, this result may only be an artifact caused by one of our assumptions, namely the fact that we consider the same disc-coupling timescale, independently of the initial mass and metallicity.

Many factors indeed affect the physics of the disc. It is not clear yet which of the photo-evaporation mechanism, accretion-related processes, or a combination of planet formation mechanism and photo-evaporation is dominant in the disc dispersal process \citep[\eg][]{Alexander2014,Gorti2015}. The in-situ planet formation process is now known to open large gaps in proto-planetary discs \citep{Alma2015} that could contribute to a more efficient disc dispersal by photo-evaporation \citep{Alexander2014b}. On one hand, if the photo-evaporation mechanism is dominant, the higher luminosity of a metal-poor star should provoke a quicker disc dispersal. But on the other hand, metallicity is a direct indicator of the condensible materials available in the disc to form planets. Planetary system formation simulations by \cite{Dawson2015} and observations from the Kepler mission show a larger fraction of large planets in metal-rich environments \citep[\eg][]{Narang2018,Cabral2018}. \cite{Mamajek2009} and \cite{Muldersbook2018} suggest that metal-rich stars would lose their disc earlier because of planet formation. 

\subsection{Rossby numbers}
\label{sect:Rossby}
The efficiency of the dynamo process that is expected to be responsible for the stellar magnetic field can be
characterized by the Rossby number defined in Eq.~\ref{Eq:Rossby}. 
The lower the Rossby number, the more active the dynamo engine, until the magnetic field eventually saturates. Given the wide range of convective envelope scales (in mass and radial extent), the depth at which the turnover timescale is computed is particularly relevant. \cite{Charbonnel2017} explored this parameter space and proposed several options that we provide in the online material as described in Table~\ref{tab:gridtable}.

\begin{center}
\begin{figure}
\includegraphics[width=0.48\textwidth]{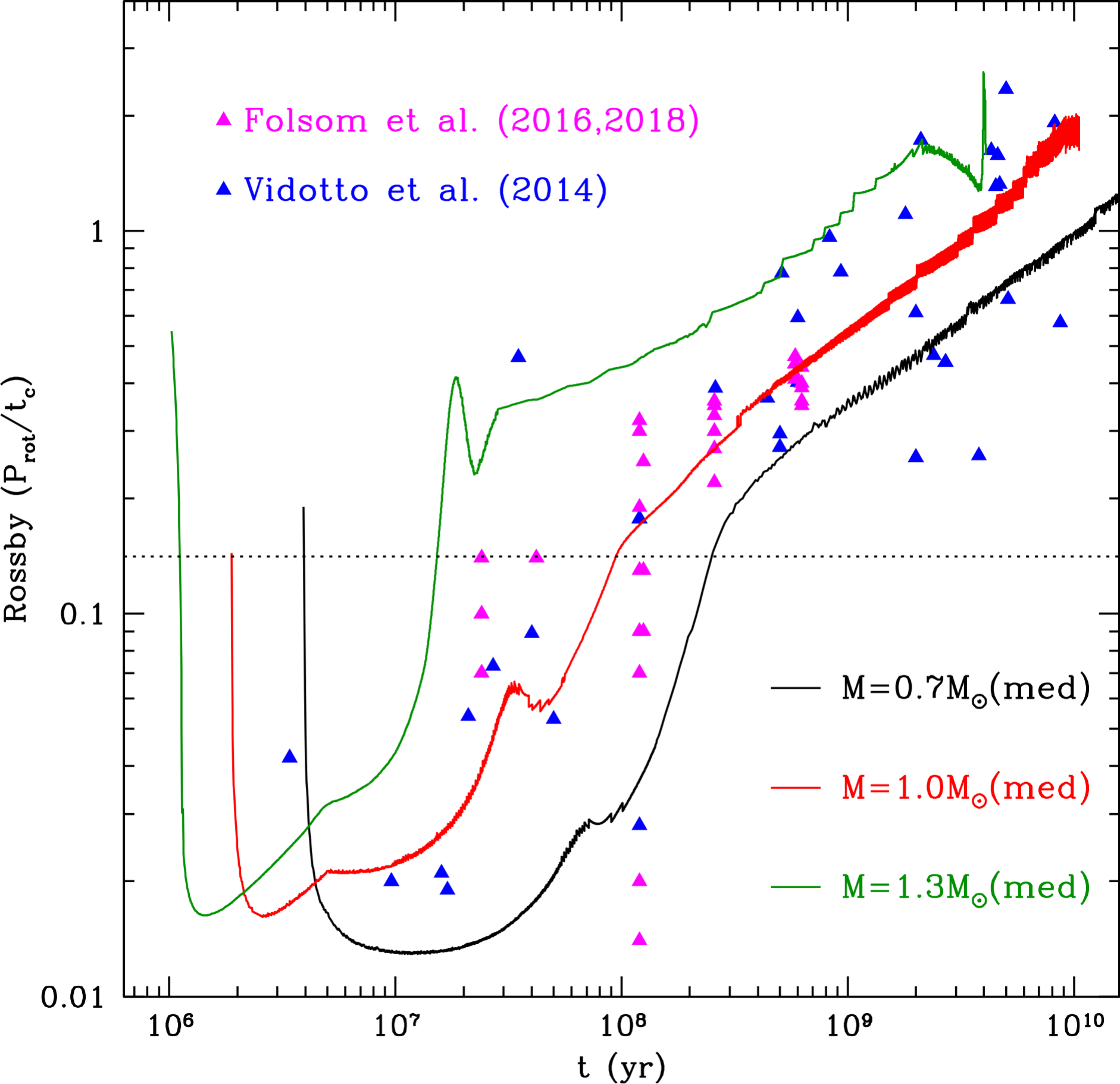}
\caption{Rossby number as a function of time for the 0.7 (black), 1.0 (red) and 1.3 (green)\Ms models for median rotators. The Rossby number is estimated at half a pressure scale height above the base of the convective envelope. The dotted line indicates the saturation Rossby number at $Ro=0.14$. The blue and magenta triangles indicates the semi-empirical Rossby numbers for solar-like stars given in \citet{Folsom2016,Folsom2018} and \citet{Vidotto2014b}.}
\label{Fig:Rossby}
\end{figure}
\end{center}

We show in Fig.~\ref{Fig:Rossby} the evolution of the Rossby number for median rotators with three different initial masses at solar metallicity compared to semi-empirical values of solar-like stars taken from the literature. In the present case, we compute the Rossby number according to Eq.~\ref{Eq:Rossby}, with the characteristic turnover timescale taken at half a pressure scale height above the base of the convective envelope. 
The Rossby number sharply drops when the radiative core appears before increasing more slowly as the envelope becomes thiner. 
Subsequently, the spin down due to magnetic winds explains the increase of the Rossby number, up to the end of the MS.
Also, the lower the stellar mass, the smaller the Rossby number due to the more extended convective envelope.

As in \citet{Charbonnel2017}, we compare our models to semi-empirical Rossby numbers taken from observational studies. We selected the observations by \citet{Folsom2016,Folsom2018} carried out as part of the ToUpiES\footnote{http://ipag.osug.fr/Anr\_Toupies/} project, and the compilation by \citet{Vidotto2014b}. They use spectro-polarimetric data to study the evolution of magnetic field with rotation and time and provide Rossby numbers that they estimated using different methods. We selected the stars with $M\in [0.7;1.3]M_\odot$, in the two samples and as can be noticed in Fig~\ref{Fig:Rossby}, our medians rotators are in good agreement with both of their samples on the PMS and the MS. 

\subsection{Lithium surface abundance}
\label{sect:Lithium}
\begin{center}
\begin{figure}
\includegraphics[width=0.48\textwidth]{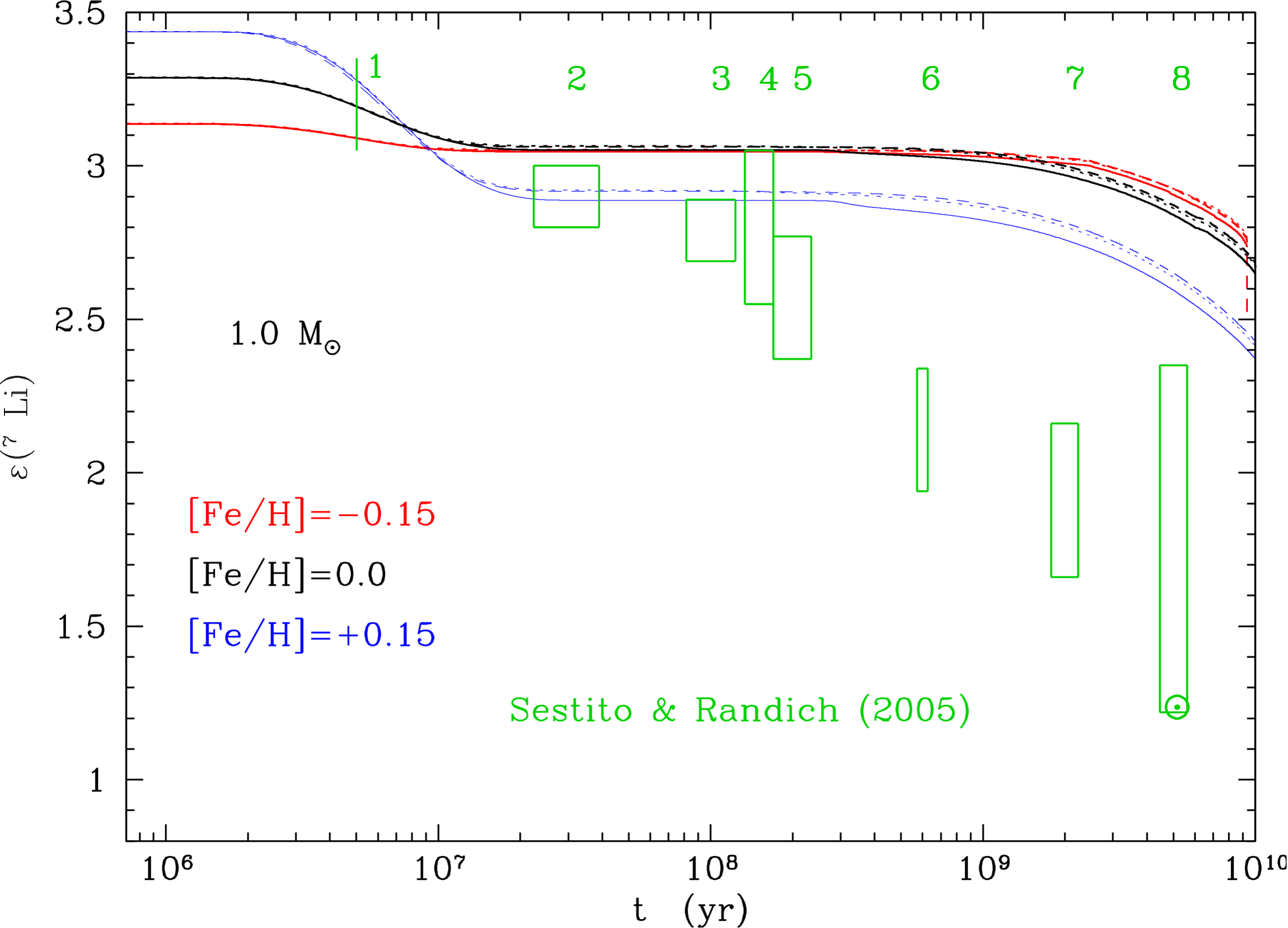}
\caption{Evolution of the $^7$Li surface abundance for our rotating 1~M$_\odot$ models at different metallicities. The solid, dotted and dashed lines refer to the fast, median and slow rotating case, respectively. We overplotted the spectroscopic $^7$Li abundances observed in some open clusters and collected by \citet{SestitoRandich2005}. The numbers 1 to 8 identify the clusters:  1) NGC2264, 2) IC2391,IC2602, and IC4665, 3) Pleiades and Blanco I, 4) NGC2516, 5) M34, M35 and NGC6475, 6) Hyades, Praesepe, Coma Ber and NGC6633, 7) NGC752, NGC36780, and IC4651, and 8) M67. The Solar abundance is indicated with $\odot$.}
\label{Fig:Lithium}
\end{figure}
\end{center}

As mentionned in the introduction, rotation induced mixing processes associated to meridional circulation and shear-induced mixing cannot explain by themselves the $^7$Li abundances observed in open clusters. 
Classically, the higher the differential rotation in the tachocline\footnote{The tachocline is the transition region between the radiative interior and the convective envelope}, the more efficient the mixing and the more important the depletion of $^7$Li. In the present case, our 1\Ms models (Fig.~\ref{Fig:Lithium}) can not reproduce the observed main sequence lithium depletion observed for t $> 10^9$ yr. Additional processes including extra mixing in the tachocline \citep[see \eg][]{Brun1999,CD2018}, or internal gravity waves \citep[\eg][]{CharbonnelTalon2005Science} are required to account for this feature.

\section{Conclusions}\label{sec:conclusion}

The present work may be considered as an update of the grid of PMS models and isochrones by \citet{Siess2000}. We presented the first grid of stellar models of low-mass PMS and MS stars including a self-consistent treatment of  the effects of rotation. 
The grid extends from 0.2\Ms up to 1.5\Ms for seven metallicities and includes  state-of-the-art micro- and macro-physics with improved surface boundary conditions and up-to-date treatment of anisotropic turbulence \citep{Mathis2018}. Our  standard  solar metallicity models are thoroughly compared with a large set of available evolutionary tracks, and besides differences on the MS of very low mass stars related to the equation of state, they show a  good agreement on the whole evolution. 
However, significant differences in terms of age and evolutionary timescale appear between different grids and are likely due to a complex interplay between the physical grid assumptions and the numerics of each code. After calibration of the solar rotation with three parameters for the braking law and none for the internal transport, 
the models are able to reproduce fairly well the evolution of the surface rotation rate observed in associations and open clusters at solar metallicity over the entire mass range 0.2-1.5\Ms. 
However, they still fail to account for the observed main sequence lithium depletion observed in the more evolved open clusters.
We also confronted our models to asteroseismic data probing the core rotation rate. We found a good agreement between our mid-F type stars models and the observations but below $\approx 1$\Ms, our models cannot explain anymore the slow core rotation rates claimed by \citet{Benomar2015}. 
 Finally, we compared our model predictions to semi-empirical Rossby numbers determinations and found a very good agreement. 
We also showed that metallicity has a strong impact on the AM losses and on the rotation period evolution of low-mass stars. Metal-poor stars are spinning faster than metal-rich ones.
We provide extended tables describing the evolution of key stellar parameters, including asterosesismic quantities and Rossby numbers. Data are available on the Geneva website\footnote{\url{https://www.unige.ch/sciences/astro/evolution/en/database/}}. They are integrated in the \textsc{Syclist} tool allowing the computation of isochrones and synthetic clusters \citep{Georgy2014b}.

The computation at different metallicities offers the possibility to compare the grid to new incoming data. Among others (TESS or PLATO), Gaia is expected to provide rotation periods and spectroscopic data for a few million stars. This can be a fantastic playground and a great opportunity to test the robustness of rotational treatment for different chemical compositions.

\begin{acknowledgements}
 This study was supported by the grant ANR 2011 Blanc SIMI5-6 020 01 ?Toupies: Towards understanding the spin evolution of stars? (http:\/\/ipag.osug.fr\/Anr\_Toupies\/ ). LA thanks the European Research Council through the grant ERC 682393 (AWESoMeStars). CC, LA, FG and CG thank the Equal Opportunity Office of the University of Geneva. LA, AP, CC, FG, and NL thank the "Programme National de Physique Stellaire" (PNPS) of CNRS/INSU co-funded by CEA and CNES for financial support. FG acknowledge financial support from the CNES fellowship. LS is senior FNRS research associate.
\end{acknowledgements}

\bibliography{BibADS}

\begin{appendix}
\section{Content of the electronic tables}
\begin{onecolumn}
\begin{longtable}{l l l}
  \caption{Description of quantities stored in the electronic tables.}\\
  \hline \hline
  \bf{Stellar parameters}                                                                               & \bf{Surface abundances}           & \bf{Central abundances}           \\
  \hline \hline
\endfirsthead
 \hline
\caption{Continued.} \\
\hline
  \bf{Stellar parameters}                                                                               & \bf{Surface abundances}           & \bf{Central abundances}           \\
\hline \hline \\
\endhead

 - Age t (yr)         																					& $^{1}$H $^{2}$H                   & $^{1}$H $^{2}$H                   \\
 - Effective temperature log(T$_{\mathrm{eff}})$ (log(K))   											& $^{3}$He $^{4}$He       	        & $^{3}$He $^{4}$He                 \\
 - Surface luminosity log(L) (log(L$_{\odot}$))     													& $^{6}$Li $^{7}$Li                 &    				                \\
 - Surface gravothermal luminosity log(L$_\textrm{grav}$) (log(L$_{\odot}$))                            & $^{7}$Be $^{9}$Be                 &                                   \\
 - Stellar mass M (\Ms)                                       				                            & $^{10}$B $^{11}$B                 &                                   \\
 - Photospheric radius R$_\mathrm{eff}$ (R$_{\odot}$)                   								        & $^{12}$C $^{13}$C $^{14}$C        & $^{12}$C $^{13}$C $^{14}$C        \\
 - Photospheric density $\rho_\mathrm{eff}$ (g.$\mathrm{cm^{-3}}$)       										& $^{14}$N $^{15}$N                 & $^{14}$N $^{15}$N                 \\
 - Photospheric gravity log($g_\mathrm{eff}$)  (log(cgs))        					                & $^{16}$O  $^{17}$O $^{18}$O       & $^{16}$O $^{17}$O $^{18}$O        \\
 - Mass loss rate (\Ms yr$^{-1}$)                                  					                    & $^{19}$F                          & $^{19}$F                          \\
 - Central temperature log($T_{c}$) (log(K))                          								    & $^{20}$Ne $^{21}$Ne $^{22}$Ne     & $^{20}$Ne $^{21}$Ne $^{22}$Ne     \\
 - Central pressure $P_{c}$                                                                             & $^{23}$Na                         & $^{23}$Na      				    \\
 - Central density $\rho_{c}$  							                                                & $^{24}$Mg $^{25}$Mg $^{26}$Mg     & $^{24}$Mg $^{25}$Mg $^{26}$Mg     \\
 - Maximum temperature $T_{\mathrm{max}}$	 (K)    												    & $^{26}$Al $^{27}$Al       	    & $^{26}$Al $^{27}$Al     	      	\\
 - Mass coordinate of $T_{\mathrm{max}}$    (\Ms)     													& $^{28}$Si   					    & $^{28}$Si              		    \\
 - Density at the location of $T_{max}$, $\rho_{max}$ (g.$\mathrm{cm^{-3}}$)                            &                                   &                                   \\
 - Central value of the total nuclear energy production rate $\varepsilon_\mathrm{nuc,c}$ ($\mathrm{erg\, g^{-1} s^{-1}}$)            	    &								   &\\
 - Central value of the gravothermal energy production rate $\varepsilon_\mathrm{grav,c}$ ($\mathrm{erg\, g^{-1} s^{-1}}$)            	    &								   &\\
 - Central value of the plasma neutrino energy loss rate $\varepsilon_\mathrm{\nu,c}$ ($\mathrm{erg\, g^{-1} s^{-1}}$)             	        &								   &\\
 &&\\     		
 - Mass at the base of CE M$_{BCE}$ (\Ms)        														&								    &								    \\
 - Radius at the base of CE R$_{BCE}$ (\Rs)        														&								    &								    \\
 - Temperature at the base of CE log(T$_{BCE}$) (log(K))        										&								    &								    \\
 - Density at the base of CE $\rho_{BCE}$ (g.$\mathrm{cm^{-3}}$)        								&								    &								    \\
 - Mass at the top of CC M$_{CC}$ (\Ms)        															&								    &								    \\
 - Radius at the top of CC R$_{CC}$ (\Rs)        														&								    &								    \\
 - Temperature at the top of CC log(T$_{CC}$) (log(K))        											&								    &								    \\
 - Density at the top of CC $\rho_{CC}$ (g.$\mathrm{cm^{-3}}$)        									&								    &	  			                    \\
&&  \textbf{Color}\\
 - Maximum convective turnover timescale in the CE $\tau_{\mathrm{conv,env,max}}$ (yr) 					&								    &                                   \\
 - Associated Rossby number $Ro_{\mathrm{env,max}}$ 													&								    &	Bolometric magnitude			\\
 - Integrated convective turnover timescale in the CE $\tau_{\mathrm{conv,env,g}}$ (yr) 				&								    &	Bolometric corrections     		\\
 - Associated Rossby number $Ro_{\mathrm{env,global}}$ 													&									&	U-B							    \\ 
 - Convective turnover timescale at $H_p/2$ above the base of the CE $\tau_{\mathrm{conv,env,Hp/2}}$ (yr) &									&	B-V								\\
 - Associated Rossby number $Ro_{\mathrm{env,Hp/2}}$ 													&									&	V-R								\\
 - Convective turnover timescale at $H_p$ above the base of the CE $\tau_{\mathrm{conv,env,Hp}}$  (yr) 	&									&	V-I								\\
 - Associated Rossby number $Ro_{\mathrm{env,Hp}}$ 														&									&	J-K								\\
 - Convective turnover timescale at mid radius CE $\tau_{\mathrm{conv,env,midRCE}}$  (yr) 				&									&	H-K								\\ 
 - Associated Rossby number $Ro_{\mathrm{env,midRCE}}$ 													&									&	V-K								\\
 - Convective turnover timescale at mid mass CE $\tau_{\mathrm{conv,env,midMCE}}$  (yr) 				&									&	G-V								\\ 
 - Associated Rossby number $Ro_{\mathrm{env,midMCE}}$ 													&									&	G$_\textrm{BP}$-V				\\
 							
 - Maximum convective turnover timescale in the CC $\tau_{\mathrm{conv,core,max}}$ (yr) 				&									&   G$_\textrm{RP}$-V               \\
 - Associated Rossby number $Ro_{\mathrm{core,max}}$ 													&									&	M$_\textrm{U}$	    			\\
 - Integrated convective turnover timescale in the CC $\tau_{\mathrm{conv,core,g}}$ (yr) 				&									&	M$_\textrm{B}$	    			\\
 - Associated Rossby number $Ro_{\mathrm{core,global}}$ 												&									&	M$_\textrm{V}$	    			\\
 - Convective turnover timescale at $H_p/2$ below the top of the CC $\tau_{\mathrm{conv,core,Hp/2}}$ (yr) &									&	M$_\textrm{R}$	    			\\
 - Associated Rossby number $Ro_{\mathrm{core,Hp/2}}$ 													&									&	M$_\textrm{I}$	    			\\
 - Convective turnover timescale at $H_p$ below the top of the CC $\tau_{\mathrm{conv,core,Hp}}$  (yr) 	&									&	M$_\textrm{H}$	    			\\
 - Associated Rossby number $Ro_{\mathrm{core,Hp}}$ 													&									&	M$_\textrm{J}$	    			\\
 - Convective turnover timescale at mid radius CC $\tau_{\mathrm{conv,core,midRCC}}$  (yr) 				&									&	M$_\textrm{K}$	    			\\ 
 - Associated Rossby number $Ro_{\mathrm{core,midRCC}}$ 												&									&	M$_\textrm{G}$	    			\\
 - Convective turnover timescale at mid mass CC $\tau_{\mathrm{conv,core,midMCC}}$  (yr) 				&									&	M$_\mathrm{G_\textrm{BP}}$	    \\ 
 - Associated Rossby number $Ro_{\mathrm{core,midMCC}}$ 												&									&	M$_\mathrm{G_\textrm{RP}}$	    \\
&&\\
 - Fractional convective radius of gyration k$^2_\mathrm{conv}$ \citep[][adimensional]{Rucinski1988} 			&								    &								    \\ 
 - Fractional radiative radius of gyration k$^2_\mathrm{rad}$ \citep[][adimensional]{Rucinski1988} 			&								    &								    \\ 
 - Surface angular velocity $\Omega_s$ (rad.s$^-1$) 												    &							        &								    \\ 
 - Radiative (+ convective) core mean angular velocity $\Omega_c$ (rad.s$^-1$) 						    &								    &								    \\ 
 - Surface velocity $v_\mathrm{surf}$ ($\mathrm{km.s^{-1}}$) 													&								    &								    \\ 
 - Surface rotation period P$_\mathrm{rot}$ (days)    															&								    &								    \\ 
 - Total specific angular momentum content of the star J$_{act}$=$\frac{1}{M_\mathrm{tot}}\int_0^{M_\textrm{tot}}\Omega r^2 dm$ (cgs)       &								   &\\
 - Angular momentum content of the core J$_{core}$=$\int_0^{M_{BCE}}\Omega r^2 dm$ (cgs) 				&								    &								    \\
 - Ratio $\Omega/\Omega_\mathrm{crit}$														            &								    &								    \\
 - Break-up surface velocity ($\mathrm{km.s^{-1}}$) 			                                        &								    &								    \\
 - Angular momentum torque at the surface from magnetized stellar winds									&								    &								    \\
 - Equipartition magnetic field according to \citet{CS2011}												&								    &								    \\
 && \\ 
 - The large separation from asymptotic relation $\Delta \nu_{\mathrm{asymp}}$ ($\mathrm{\nu Hz}$) 		& 							    	&	 							    \\
 - The large separation from scaling relation $\Delta \nu_{\mathrm{scale}}$ ($\mathrm{\nu Hz}$) 		& 								    & 							      	\\
 - Relative error on large separation $\frac{(\Delta \nu_{\mathrm{asymp}}-\Delta \nu_{\mathrm{scale}})}{\Delta \nu_{\mathrm{asymp}}}$ $\Delta \nu_{\mathrm{err}}$		      &&\\
 - The frequency with the maximum amplitude $\nu_{\mathrm{max}}$										& 								    &								    \\
 - Asymptotic period spacing of g-modes $\Delta\mathrm{\Pi}$ (s)										& 								    & 							      	\\ 
 - The total acoustic radius T (s) 																		& 								    & 							      	\\  
 - Acoustic radius at the base of the convective envelope t$_{\mathrm{BCE}}$ (s)  						& 								    & 							      	\\  
 - Acoustic radius in the helium second-ionisation region t$_{\mathrm{He}}$ (s)							& 								    & 							      	\\  
 && \\ 	

   \label{tab:gridtable}
\end{longtable}
\end{onecolumn}
\end{appendix}
\end{document}